\documentclass[12pt]{article}
\pdfoutput=1

\usepackage[dvipdfmx]{graphicx}
\usepackage{graphicx,bm,color}
\usepackage{amsmath}
\usepackage{amssymb}
\usepackage{amsfonts}
\usepackage{cases}
\usepackage{cancel}
\usepackage{hyperref}
\usepackage{ulem}
\usepackage{cite}
\usepackage{booktabs}
\usepackage{multirow}
\usepackage{slashed}

\newcommand{\gsim}{\lower.7ex\hbox{$\;\stackrel{\textstyle>}{\sim}\;$}}
\newcommand{\lsim}{\lower.7ex\hbox{$\;\stackrel{\textstyle<}{\sim}\;$}}

\begin{document}
	
	\vspace*{8mm}
	
	\begin{center}
		
		{\Large\bf  On the phenomenological implications of a Pati-Salam model spontaneously broken by Higgs fields in fundamental representations }
		
		\vspace*{9mm}
		
		\mbox{\sc George K. Leontaris$^{1}$, Ruiwen Ouyang$^2$,  Ye-Ling Zhou$^2$ }\vspace*{3mm}

		{\small
			
			$^1$  Physics Department, University of Ioannina, University Campus, Ioannina 45110, Greece. \\ \vspace{0.3cm}
			
			$^2$ School of Fundamental Physics and Mathematical Sciences,\\ Hangzhou Institute for Advanced Study, UCAS, Hangzhou 310024, China. \\ \vspace*{0.3cm}
			
		}
		
	\end{center}
	
	\vspace{20pt}
	
	\begin{abstract}
		
		\noindent
		We present a supersymmetric Pati-Salam model with small representations as a potential candidate for physics beyond the Standard Model. The model features a Higgs sector with bifundamental fields $H_R+\bar H_R=(4,1,2)+(\bar 4,1,2)$, $H_L+\bar H_L=(4,2,1)+(\bar 4,2,1)$ as well as a pair of bi-doublet fields $h_a=(1,2,2)$ where $ a=1,2$, with three families of fermions accommodated in $ (4,2,1)+(\bar 4,1,2)$ as usual. The matter spectrum is augmented with three copies of neutral singlets that mix with ordinary neutrinos to realize the seesaw mechanism. The model introduces supersymmetric R-symmetry and a global discrete $\mathbb{Z}_n$ symmetry ($n > 2$) that prevents disastrous superpotential couplings, while its spontaneous breaking implies the existence of domain walls that are successfully addressed. Interestingly, the one-loop beta coefficient of the $SU(4)_C$ gauge coupling is zero in the minimal $\mathbb{Z}_3$ model, rendering the corresponding gauge coupling near-conformal in the UV. Meanwhile, Landau poles are avoided up to the Planck scale and proton decay is suppressed, resulting in a proton lifetime beyond current experimental bounds. By virtue of the extended Higgs sector, the key advantage of this PS model is its ability to disentangle quark and lepton masses through higher-dimensional effective operators, addressing a common limitation in GUT models with small Higgs representations. This makes the model more economical and easier to be constructed from string theory, particularly in several heterotic and F-theory models where Higgses in the adjoint representation are absent.

	\end{abstract}
	
	\clearpage
	
\section{Introduction}
	
	The Pati-Salam (PS) model is based on the $G_{PS}=SU(4)_C \times SU(2)_L \times SU(2)_R $ gauge symmetry~\cite{Pati:1974yy}, with the leptons and quarks of each generation grouped together in a single representation, the fourplet of $SU(4)_C$. The $G_{PS}$ gauge group is the simplest extension of the Standard Model(SM) which provides a robust framework for a unified theory. It is a left-right symmetric model, with left-handed (LH) fermionic fields accommodated in $F_L^i = (4,2,1)$ under $G_{PS}$ gauge factors, and right-handed (RH) ones in $\bar{F}_R^i = (\bar{4},1,2)$ where the index refers to the families, $(i=1,2,3)$. Remarkably, the latter representation, in addition to SM RH-fermions, incorporates the right-handed neutrinos in a natural manner. The supersymmetric version of the Standard Model can be readily obtained as a low-energy effective theory after the spontaneous breaking of the Pati-Salam symmetry through the vacuum expectation values (VEVs) of appropriate Higgs fields.

	The	$G_{PS}$ gauge symmetry also arises as an intermediate step between a fully unified theory such as $SO(10)$ or $E_6$ and the SM gauge group. Compared to the $SU(5)$ Grand Unified Theory (GUT)~\cite{Georgi}, it exhibits some advantages which are worth mentioning. For instance,  rapid proton decay which is a notoriously challenging issue at least in the minimal $SU(5)$, can be naturally suppressed in the PS constructions, making model building more attractive for low energy phenomenological explorations. Also, the $G_{PS}$ symmetry can break to the SM gauge group by only employing small Higgs representations, e.g, in the fundamental of $SU(4)_C$, unlike the $SU(5)$ which requires Higgs fields in the adjoint representation. By virtue of this property, in recent decades, the $G_{PS}$ gauge symmetry has been intensively studied as an effective model derived from the various limits of string theory.  
	The supersymmetric Pati-Salam model was initially proposed as a possible string candidate model in~\cite{Antoniadis:1990hb} and was further explored~\cite{heterotic} in the context of heterotic string model building, where in this particular context adjoint representations are not available to accommodate the Higgs fields associated with the spontaneous breaking of $SU(4)_C$.
	Subsequent studies of PS models derived with intersecting D-branes~\cite{intersectingbrane} and in F-theory constructions~\cite{Beasley:2008dc} yielded new versions with variant spectra and alternative symmetry breaking mechanisms, thus revealing several novel features.
	F-theory, in particular, is very promising for building particle physics models that reproduce the known phenomenology compatible with current experiments and suggest testable predictions.

	The minimal version~\cite{Antoniadis:1988cm} of the supersymmetric PS model entails three pairs of $F_L=(4,2,1)$ and $\bar F_R=(\bar 4,1,2)$ representations to accommodate the three fermion generations. The Higgs sector comprises of a pair residing in $H=(4,1,2)$ and $\bar H=(\bar 4,1,2)$ which breaks the PS symmetry $SU(4)_C \times SU(2)_R\to SU(3)_C\times U(1)_{B-L}$ and a Higgs bi-doublet $h=(1,2,2)$. The latter decomposes to the two MSSM Higgses,  which break the SM gauge group and give masses to quarks and leptons. Furthermore, when a sextet $D=(6,1,1)$ is included it forms superpotential couplings with $H,\bar H$ which generate heavy masses for the down-type color triplets in $H,\bar H$ Higgses left `uneaten' by the superhiggs mechanism. Having achieved this, the massless spectrum left in the model includes only the MSSM fields enhanced by three copies of RH neutrinos.

	However, although after symmetry breaking all fermions acquire non-zero masses, the model in its minimal version described above fails to reproduce the correct fermion mass spectrum and the mixing pattern observed at low energies. The main reason is that all masses of the chiral fermions arise from a single term of superpotential $W$, namely $\bar F_L F_R h \subset W$, which predicts that the down-quark and charged lepton mass matrices are equal, $m_d=m_{\ell}$, at the PS scale. If no other contributions are taken into account, then, as is well known their evolution down to low energies due to renormalization group running gives predictions incompatible with the experimentally measured fermion mass ratios, especially for the lighter generations.
	Possible ways out of the impasse have been suggested based on the inclusion of the $SU(4)_C$ adjoint  (see for example~\cite{Anastasopoulos}) and higher order corrections, however, there are two possible issues in this approach. Firstly, introducing large representations such as the adjoint ${\bf 15}\in SU(4)_C$ might lead to a Landau pole well below the Planck scale due to the rapid growing of the gauge couplings. Secondly, as noted earlier, several appealing string derived constructions do not contain the adjoint representation and therefore, the associated benefits cannot be reaped.  On the contrary, most of the PS models built in the context of the various limits of string theories do predict an excess of $SU(4)_C \times SU(2)_{L}\times SU(2)_R$ bifundamental representations  $(4,2,1)+(\bar 4,2,1 )$ where additional Higgs fields can be accommodated.

	Therefore, in this study, we draw inspiration from superstring realizations of the Pati-Salam symmetry and leverage the presence of Higgs fields in bifundamental representations $(4,2,1)+(\bar 4,2,1)$ with respect to $SU(4)_C\times SU(2)_L\times SU(2)_R$ symmetry. When these Higgs fields acquire vevs, they play a crucial role in breaking the equality of quark-lepton mass matrices which are typically predicted by GUTs. By extending previous Pati-Salam models through the inclusion of Higgs fields in bifundamentals, we demonstrate the ability to distinguish the down quark mass matrix from the charged lepton mass matrix. This distinction provides a unique opportunity to make accurate predictions of mass ratios at the electroweak scale.

	In this work, we adopt a bottom-up approach to construct a supersymmetric Pati-Salam model extended by a global $U(1)_R$ symmetry and a discrete $\mathbb{Z}_n$ symmetry under which the superfields are charged, with $n=3$ chosen due to minimality. These global symmetries play a crucial role in preventing disastrous terms in the superpotential, ensuring the stability and consistency of the model. We will show that the GUT prediction of the down quark-lepton mass matrix equality is modified after introducing appropriate bifundamental Higgs fields as discussed above, whose vevs effectively split the mass matrices, paving the way for accurate predictions of low-energy fermion mass relations. Meanwhile, except the SM degrees of freedom, all other fields are decoupled from the low energy theory either because they are too heavy, or due to the absence of direct couplings with the SM fields without further assumptions. 
	Additionally, in the minimal setup, the runnings of gauge couplings are investigated which shows that the $SU(4)_C$ couplings become near-conformal after the PS scale. We also analyze proton decay within this framework and find that the relevant contributions are significantly suppressed, resulting in a long proton lifetime that exceeds current experimental limits. Furthermore, we explore the implications of spontaneous breaking of the $\mathbb{Z}_3$ symmetry, particularly the formation of topological defects such as domain walls, with carefully adjusted energy scales, and we investigate the collapse of these domain walls and the conditions under which they can produce detectable gravitational waves. This analysis provides insights into the cosmological consequences of the model and opens up the possibility of connecting high-energy physics with observable gravitational wave signatures. Overall, our work presents a comprehensive framework that bridges particle physics, cosmology, and astrophysics, offering testable predictions and addressing key theoretical challenges.

    The layout of the paper is as follows. In section~\ref{sec:2} we describe the basic features of the model, including its symmetries, the particle content, and the superpotential. We also discussed with great details about the splitting masses between the down quark and the charged lepton mass matrices, the mechanism giving rise to masses of neutrinos, and the decoupling of heavy triplets and heavy doublets. In section~\ref{sec:3}, we analyze the renormalization group running of gauge couplings from the electroweak scale to the GUT scale. Without assuming the $SO(10)$ unification, the $SU(4)_C$ gauge coupling becomes near-conformal and the rest of the gauge couplings are safe from developing a Landau pole up to the Planck scale, rendering the model UV-consistent. In section~\ref{sec:proton-decay}, it is shown that proton decay is naturally suppressed in the proposed PS model. In section~\ref{sec:5}, we discuss the cosmological implications focusing on the domain wall problem which appears due to the spontaneous breaking of the discrete $\mathbb{Z}_3$ symmetry. In section~\ref{sec:6} we present our conclusions.

\section{A Pati-Salam model with bifundamental Higgses}\label{sec:2}
	In the supersymmetric Pati-Salam model~\cite{Antoniadis:1988cm,Antoniadis:1990hb}, all SM quarks and leptons as well as three families of right-handed neutrinos are unified into three copies of chiral superfields $F_L^i = (4, 2, 1)$ and $\bar{F}^j_R = (\bar{4}, 1, 2)$  ($i,j=1,2,3$) charged under the gauge symmetry $SU(4)_C \times SU(2)_L \times SU(2)_R$. Specifically, for each generation, these SM chiral fermions are assigned into the representation 
	\begin{eqnarray}
		&&
		F_L^i = \begin{pmatrix}
			u_1 & u_2 & u_3 & \nu \\
			d_1 & d_2 & d_3 & e
		\end{pmatrix}^i \, ,  \,
		\bar{F}_R^j =  \begin{pmatrix}
			d_1^c & d_2^c & d_3^c & e^c \\
			u_1^c &u_2^c &u_3^c & \nu^c  
		\end{pmatrix}^j \, ,
	\end{eqnarray}
	where the subscripts of $u$ and $d$ refer to the color indices, and the notation is used following the reference~\cite{Allanach:1996hz}. The SM Higgs bosons are contained in (a linear combination of) left-right bi-doublet representations $h^a = (1,2,2)$ which in general have the form
	\begin{eqnarray}
		h^a = \begin{pmatrix}
			h_u^+ & h_d^0 \\
			h^{0}_u & h^{-}_d
		\end{pmatrix}^a \, . 
	\end{eqnarray}
    The doubling of the Higgs bi-doublets $a=1,2$ is crucial to the number of light Higgs bosons with electroweak-scale masses. Indeed, in sections \ref{sec:mass-matrices}, we will show that the minimal case with $a=1$ can hardly reproduce correct fermion mass hierarchies at low energy, so we need at least $a=2$. Then, in section \ref{sec:HiggsDoublets}, a seesaw-like mechanism is applied to Higgs bosons so that only two combinations become light to form $H_u$ and $H_d$ in MSSM.

	In addition to SM fermions and Higgses, the PS model must contain additional multiplets to spontaneously break the PS symmetry down to the SM gauge group. In the supersymmetric context~\cite{Antoniadis:1988cm}, this can happen at the PS scale $v_{PS} \gg v_{EW}$ when the following Higgs bifundamentals are introduced: 
	\begin{eqnarray}
		&&
		{H}_R =(4,1,2)=  \begin{pmatrix}
			\bar d_H^c & \bar d_H^c & \bar d_H^c & \bar e_H^c \\
			\bar u_H^c &\bar u_H^c &\bar u_H^c &\bar\nu_H^c  
		\end{pmatrix} \, ,  \,
		\bar{H}_R = (\bar 4,1,2)=  \begin{pmatrix}
			d_H^c & d_H^c & d_H^c & e_H^c \\
			u_H^c &u_H^c &u_H^c & \nu_H^c  
		\end{pmatrix} \, .
	\end{eqnarray}
	Notice that $H_R$ and $\bar{H}_R$ are both $SU(2)_R$ doublets and they acquire vevs along their neutral directions at 
	$v_R \simeq \bar{v}_R \simeq v_{PS}$, where 
	\[ v_R =\langle \nu_H^c \rangle,\;\;  \bar v_R =\langle \bar \nu_H^c \rangle~.\]
	As mentioned in the introduction, the minimal version of this model predicts 
	unification of fermion masses at the PS scale $m_d = m_\ell$ for all three generations, and as is well-known in this case the low energy observed data is unlikely to be reproduced.
	
	In this paper, motivated by typical massless spectra that appear in string-derived models, we also include pairs of bifundamentals $H_L=(4,2,1)$ and $\bar{H}_L=(\bar{4},2,1)$ which transform non-trivially under $SU(2)_L$. 
	When these Higgs fields acquire vevs, they generate sizeable corrections to the fermion mass matrices that do not maintain the above equality and create the conditions for a correct mass ratio at low energies. 
	
	As with many effective models, it is important to invoke global symmetries in order to constrain the superpotential of the model. The origin of such symmetries varies, depending on the parent theory at the ultraviolet completion limit. Among other possible symmetries, the global $U(1)_R$ symmetry~\cite{SUSY} transforms a chiral superfield as $\Phi(y,\theta) \to e^{in \epsilon} \Phi(y,e^{-i\epsilon} \theta)$, where $n\equiv R(\Phi)$ is the R-charge of the superfield $\Phi$. The invariance of superpotential requires that $R(W)=2$, and in case that R-symmetry is broken to R-parity, this selection rule is modified to $R(W)=2 \ {\rm mod} \ 2$. Phenomenologically interesting assignments for R-charges require the matter fields to have one unit of R-charge, and the Higgs fields to carry zero R-charge. The detailed R-charge assignments for superfields in the proposed PS model are summarized in Table~\ref{tab:superfield}.

	As was originally discussed in~\cite{Antoniadis:1988cm}, three gauge-singlets $\psi^m$ ($m=1,2,3$) can be introduced to trigger the inverse seesaw mechanism to account for the smallness of the observed neutrino masses. The fields $\psi^m$ will acquire Majorana masses when another gauge-singlet $\phi$  acquires a large vev $\langle\phi\rangle\sim m_\psi$ via the trilinear coupling  $\langle \phi \rangle \psi^m \psi^n $\footnote{Though a more complicated scenario involving non-singlets can be considered, the introduction of gauge-singlets stands out as the simplest way to generate desired mass hierarchies in the fermion sector.}. The presence of this operator thus requires that the fields $\psi^m$ carry one unit of R-charge while $\phi$ should have zero R-charge. On the other hand, though being neutral under gauge symmetries, the field $\phi$ must be non-trivially charged under a discrete symmetry which is assumed to be a typical $\mathbb{Z}_n$, to avoid having massless Goldstone bosons after the spontaneous symmetry breaking~\footnote{It is also natural to have them because string-inspired models usually come with discrete symmetries like an $\mathbb{Z}_n$ and include many globally-charged gauge-singlets in their spectrum.}. The inclusion of singlets automatically introduces an intermediate scale $\langle\phi\rangle \equiv v_\phi < v_{PS}$ that is crucial to the creation of hierarchies, and it is convenient to parametrize the ratio of scales by $r  \equiv v_\phi/v_R \ll 1 $ (in our analysis, we adopt $r< 10^{-1}$ for concreteness). The overall $\mathbb{Z}_n$ charge assignments are summarized in Table~\ref{tab:superfield}.

	Finally, the model involves $SU(4)_C$ sextets $(6,1,1)$ that couple to pairs of fourplets following the rule $(4\times 4)\times 6 =(6+10)\times 6\to 1+\cdots$. In the minimal case, one sextet suffices to form heavy mass terms with down-type color triplets $d_H^c, \bar d_H^c$ left `uneaten' by the Higgs mechanism, however, string-derived models usually predict more than one such fields. In intersecting D-brane constructions these may arise in pairs, and might be distinguished by discrete symmetries and various $U(1)$ charges~\cite{intersectingDbranes,Anastasopoulos} including the possibility of the two distinct orientations of the strings with their ends attached on the D-brane stacks. 
	
	In the present bottom-up construction, we assume that there are two types of sextets having either one unit of R-charge denoted as $T^k$ ($k=1,..., N_{T}$), or zero R-charge denoted as $D^l$ ($l=1,..., N_{D}$). In the next subsection, we will show that having these two sextets makes it possible to create a suitable arrangement for correct SM fermion mass matrices and, in the meantime, make the other non-SM fields heavy and decoupled.  {Therefore, if any one of the sextets were absent, either we can not get the mass matrices correctly because the coupling to fermions such as $F_L H_L T$ cannot be formed, or we will obtain massless color triplets in the Higgses $H_R, \bar{H}_R$ since the coupling to the Higgses such as $H_R \bar{H}_R D$ will be gone.} For this reason, we must keep the minimal number of $N_T$ and $N_D$ to be one instead of zero.

	\begin{table}[ht!]
		\renewcommand{\arraystretch}{1.5}
		\centering
		\begin{tabular}{|c|c|c|c|c|}
			\hline
			Superfields & PS reps  & SM decomposition & $U(1)_R$ & $\mathbb{Z}_n$ charge   \\
			\hline
			$F_L^{i=1,2,3}$ & $(4,2,1) $  & $Q(3,2,\frac16)+\ell(1,2,-\frac12)$ & 1 & $n_L$  \\
			\hline
			\multirow{ 2}{*}{ $\bar{F}_R^{j=1,2,3}$} & \multirow{ 2}{*}{$(\bar{4},1,2)$} & $u^c(\bar{3},1,-\frac23)+d^c(\bar{3},1,\frac13)$ & \multirow{ 2}{*}{1} & \multirow{ 2}{*}{$n_R$} \\
			&   & $+\ e^c(1,1,1)+\nu^c(1,1,0)$ &  &  \\
			\hline
			$h^{a=1,2}$ & $ (1,2,2)$ & $h^a_u(1,2,\frac12)+h^a_d(1,2,-\frac12)$ & 0 & $n_h$  \\
			\hline
			$H_L $ & $(4,2,1)$ & $Q_{H_L}(3,2,\frac16)+\ell_{H_L}(1,2,-\frac12)$ &0 & $n_{H_L}$  \\
			\hline
			\multirow{ 2}{*}{$ \bar{H}_R$} & \multirow{ 2}{*}{$(\bar{4},1,2)$}   & $u_{H_R}^c(\bar{3},1,-\frac23)+d_{H_R}^c(\bar{3},1,\frac13)$ &  \multirow{ 2}{*}{0} &  \multirow{ 2}{*}{$n_{\bar{H}_R}$}  \\
			&    & $+ \, e_{H_R}^c(1,1,1)+\nu_{H_R}^c(1,1,0)$ &  &   \\
			\hline
			$\bar{H}_L$ & $(\bar{4},2,1)$ & $\bar{Q}_{H_L}(\bar{3},2,-\frac16)+\bar{\ell}_{H_L}(1,2,\frac12)$ & 0 & $n_{\bar{H}_L}$ \\
			\hline
			\multirow{ 2}{*}{$ H_R$} & \multirow{ 2}{*}{$({4},1,2)$} & $\bar{u}^c_{H_R} ({3},1,\frac23)+ \bar{d}^c_{H_R} ({3},1,-\frac13) $ & \multirow{ 2}{*}{0} &  \multirow{ 2}{*}{$n_{H_R}$}  \\
			&  & $ + \, \bar{e}^c_{H_R} (1,1,-1) +\bar{\nu}^c_{H_R} (1,1,0) $ &  & \\
			\hline
			$\phi$ & $(1,1,1)$ & $\phi(1,1,0)$ & 0 & $n_\phi$  \\      
			\hline
			$\psi^{m=1,...,N_\psi} $  & $ (1,1,1)$ & $\psi^m (1,1,0)$ & 1  & $n_\psi$  \\
			\hline
			$T^{k=1,...,N_{T}}$ & $(6,1,1)$ & $T_3(3,1,-\frac13)+\bar{T}_3(\bar{3},1,\frac13)$ & 1 & $n_{T}$ \\ 
			\hline
			$D^{l=1,...,N_{D}}$ & $(6,1,1)$ & $D_3(3,1,-\frac13)+\bar{D}_3(\bar{3},1,\frac13)$ & 0 & $n_{D}$ \\ 
			\hline
		\end{tabular}
		\caption{Spectrum of the proposed Pati-Salam model. Additional $\mathbb{Z}_n$ charges were assigned to every superfield. A superfield $\Phi$ with $\mathbb{Z}_n$ charge $m$ transforms as $\Phi \to \omega \Phi$ ($\omega = \exp{\frac{2i m \pi}{n}}$) under the discrete $\mathbb{Z}_n$ symmetry.}
		\label{tab:superfield}
	\end{table}

	In Table~\ref{tab:superfield}, we summarize all the representations of the aforementioned superfields under the PS gauge group and their decomposition under the SM gauge group. Note that the above construction involves only the small representations, without using the $SU(4)_C$ adjoint representation $(15,1,1)$ or the $SU(2)_R$ triplet $(10,1,3)$ that are commonly used in other similar Pati-Salam constructions. This is a key feature of the proposed model. 

	In summary, in this section, we reviewed the supersymmetric PS model as a possible candidate for BSM physics at a high energy scale. After the spontaneous breaking of the PS symmetry, only SM fermions and SM Higgses stay light, so the model reduces to the conventional MSSM which is assumed to be effective from a few orders of magnitude higher than the electroweak scale. It is also assumed that a global R-symmetry and a discrete symmetry of $\mathbb{Z}_n$ (with $n \geq 3$) are present to constrain the PS model. The spectrum of the proposed model is inspired by earlier works~\cite{intersectingbrane,Beasley:2008dc,Antoniadis:1988cm,intersectingDbranes,Anastasopoulos} about top-down constructions of particle physics models from string theory, and the representations as well as the charge assignments are listed in detail in Table~\ref{tab:superfield}.

\subsection{Conditions for discrete charges}
	Before proceeding into the details of the model, it is better to first constrain the possible choices of $\mathbb{Z}_n$ charges by requiring that the $U(1)_R$-invariant superpotential at least contains the following Yukawa terms to achieve the correct mass matrices for quarks and leptons:
	\begin{eqnarray}
		W_Y \supset 
		y_a^{ij} F_L^i \bar{F}_R^j h^a +
		y_L^{ik} F_L^i H_L T^k    + 
		y_R^{jk} \bar{F}_R^j \bar{H}_R  T^k    +
		y_\psi^{jm} \bar{F}_R^j  H_R \psi^m  + 
		y_{\phi}^{mn} \psi^m \psi^n  \phi \, .
		\label{eq:Yukawa}
	\end{eqnarray}

	The first term is a usual Yukawa interaction in the PS model that generates the Dirac masses of quarks and leptons after EWSB, whose $\mathbb{Z}_n$ charges satisfy
	\begin{eqnarray}
		n_L + n_R + n_h  = 0 \ {\rm mod} \ n
		\label{eq:cc1}
	\end{eqnarray}
	An economical solution for this condition is achieved if all SM fermions and Higgses do not carry any $\mathbb{Z}_n$ charges, i.e. $n_L = n_R = n_h = 0$. We will restrict our analysis to this assumption in the rest of the paper.

	The second and third terms $F_L^i H_L T^k +  \bar{F}_R^j \bar{H}_R T^k$ are used to generate a splitting between down-type quarks and charged leptons in section~\ref{sec:mass-matrices}. These terms will remain in the superpotential  if the charges of the representations involved in these terms meet the conditions
	\begin{eqnarray}
		n_L + n_{H_L} + n_T = 0 \ {\rm mod} \ n  \, , \nonumber \\
		n_R + n_{\bar{H}_R} + n_T = 0 \ {\rm mod} \ n  \, .
		\label{eq:cc2}
	\end{eqnarray}
	Thus when $n_L = n_R  = 0$, there must be $n_{H_L} = n_{\bar{H}_R}$.

	The fourth term in eq.~(\ref{eq:Yukawa}) is required to generate heavy masses for right-handed neutrinos by the inverse seesaw mechanism in section \ref{sec:neutrino}, which implies:
	\begin{eqnarray}
		n_R + n_{H_R} + n_\psi= 0  \ {\rm mod} \ n \, .
		\label{eq:cc3}
	\end{eqnarray}
	Therefore, if $n_\psi \neq 0 $, the field $H_R$ will carry a non-zero $\mathbb{Z}_n$ charge $n_{H_R}\neq 0$ and breaks the $\mathbb{Z}_n$ symmetry together with the vev of $\phi$.
	
	Finally, to give heavy masses to gauge-singlets $\psi$, the operator $ \phi \psi^m \psi^n$ is included which implies the constraint 
	\begin{eqnarray}
		2n_\psi + n_\phi = 0  \ {\rm mod} \ n\, .
		\label{eq:cc4}
	\end{eqnarray}
	Since as we mentioned before $n_\phi$ should be non-zero to break the global $\mathbb{Z}_n$ symmetry, we can conclude that the last condition does not admit a solution for the minimal $\mathbb{Z}_2$ case whatever value $n_\psi$ takes, and therefore this condition requires $n\geq 3$.

	\subsection{A minimal $\mathbb{Z}_3$ model}
	In this section, we will discuss a minimal construction with $n=3$ based on our previous considerations. Assuming now a $\mathbb{Z}_3$ symmetry, to know the charges for other superfields, we can first fix an arbitrary value of $n_\phi$, and then solve for the conditions given in eqs.~(\ref{eq:cc2})-(\ref{eq:cc4}). Without loss of generality, it is convenient to take $n_\phi=1$ which gives $n_\psi =1$ from eq.~(\ref{eq:cc4}), and thus $n_{H_R}=2$ from eq.~(\ref{eq:cc3}). The discrete charge of the other Higgs fourplets should be determined together with the mechanism that generates heavy masses for them, which will be discussed in detail in the section \ref{sec:HiggsDTS}. For short, a condition in eq.~(\ref{eq:cc5}) is applied which requires that $n_{H_R}=n_{\bar{H}_R}$ and fixes the values of $n_D$. For simplicity, in the following we only consider the minimal case where the number of sextets is fixed to $N_T=N_D =1$. With all the above setup, the charge assignments for all the superfields are all determined except for $\bar{H}_L$, which are listed specifically in Table~\ref{fields-Z3}.
	\begin{table}[ht!]
		\renewcommand{\arraystretch}{1.25}
		\centering
		\begin{tabular}{|c|c|c|c||c|}
			\hline
			Notations & {Superfields} & {PS reps} & {$U(1)_R$} & $\mathbb{Z}_3$ charge \\
			\hline
			\multirow{3}{*}{SM fields} & $F_L^{i=1,2,3}$ & $(4,2,1) $ & 1 &   0 \\
			& $\bar{F}_R^{j=1,2,3}$ & $(\bar{4},1,2) $ & 1 &   0 \\
			& $h^{a=1,2}$ & $ (1,2,2)$ & 0 &  0 \\
			\hline
			\multirow{2}{*}{Sextets} & $T$ & $(6,1,1)$ & 1 &  1 \\ 
			& $D$ & $(6,1,1)$ & 0 &  2 \\ 
			\hline
			\multirow{4}{*}{Higgs fourplets} & $H_L $ & $(4,2,1)$ & 0 & 2 \\
			&  $ \bar{H}_R$ & $(\bar{4},1,2)$ & 0  &  2 \\
			& $\bar{H}_L$ & $(\bar{4},2,1)$ & 0 &  2  \\
			& $H_R$ & $(4,1,2)$ & 0 & 2 \\ 
			\hline
			\multirow{2}{*}{Singlets} & $\phi $  & $ (1,1,1)$  & 0 &  1 \\
			& $\psi^{m=1,...,N_\psi} $  & $ (1,1,1)$  & 1 &  1 \\
			\hline
		\end{tabular} 
		\label{fields-Z3}
		\caption{Superfields with $\mathbb{Z}_3$ symmetry. All $\mathbb{Z}_3$ charge can be determined from conditions in eqs.~(\ref{eq:cc2})-(\ref{eq:cc4}) and eq.~(\ref{eq:cc5}) after assuming the values of $n_\phi$ and $n_{\bar{H}_L}$.}
	\end{table}

	There are a few possibilities for the choice of $\mathbb{Z}_3$ charge of $\bar{H}_L$, however, we will limit our analysis to the most convenient scenario where all Higgs fourplets carrying the same discrete charges, which is compatible with a possible incorporation of the pair $ H_R$  into a single representation of a higher symmetry such as the $\overline{16}\in SO(10)$.
	
	As a result, at the renormalizable level, the most general superpotential is given by
	\begin{eqnarray}
		W &=&   W_F + W_D + W_\phi +  W_{h} \, ,
	\end{eqnarray}
	where
	\begin{eqnarray}
		W_F &=& y_a^{ij} F_L^i \bar{F}_R^j h^a +
		y_L^{i} F_L^i H_L T   + 
		y_R^{j} \bar{F}_R^j \bar{H}_R  T    +
		y_\psi^{jm} \bar{F}_R^j H_R \psi^m \,  +
		\lambda_\psi^{im} F_L \bar{H}_L \psi^m
		\nonumber\\
		W_{D} &=& \lambda_{H_L} D H_L H_L  
		+ \lambda_{\bar{H}_L} D \bar{H}_L \bar{H}_L  
		+ \lambda_{H_R} D H_R H_R
		+ \lambda_{\bar{H}_R} D \bar{H}_R \bar{H}_R  \, , \nonumber\\
		W_{\phi} &=& y_{\phi}^{mn} \psi^m \psi^n  \phi  + y_{T} T T \phi + \lambda_\phi \phi^3\, , \nonumber\\
		W_h &=&  \mu_{1}h_1^2 + \mu_{2}h_2^2 + \mu_{12}h_1h_2  \, .
		\label{eq:superpotential}
	\end{eqnarray}

        \noindent
	As will be discussed in the following subsections, the superpotential $W_F$ generates attainable mass matrices for quarks and leptons, while terms in $W_D$ ensure that all Higgs fourplets as well as the sextet $D$ acquire heavy masses at the PS scale $v_R$, and $W_\phi$ give masses to the singlets $\psi$ and the sextet $T$ at an intermediate scale $v_\phi$. {The last term  $W_h$ is the common supersymmetric $\mu$-term triggering the EWSB, acting as a bare mass term which can be eliminated by imposing a higher symmetry at a UV scale such as the classical scale invariance. Though absent at tree-level, the $\mu$-term automatically arises either through the Giudice-Masiero mechanism~\cite{Giudice:1988yz} or by higher dimensional operators including the gauge-singlet $\phi$ acquiring vev at $v_\phi$.}

	\subsection{Quarks and lepton mass matrices}
	\label{sec:mass-matrices}
	The Higgs spectrum adopted in this model may offer a potential solution to the fermion mass problem created by
	 the tree-level Yukawa coupling relation at the PS scale.	To see how the Yukawa interactions given in eq.~(\ref{eq:Yukawa}) lead to the correct mass matrices for quarks and leptons, we need to expand each of the terms with their SM decompositions, and then find out the effective mass terms corresponding to quarks and leptons after symmetry breaking. Starting from the first term in eq.~(\ref{eq:Yukawa}), the expansion of $F_L \bar{F}_R h^a$ gives:
	\begin{eqnarray}
		y_a^{ij} F_L^i \bar{F}_R^j h^a =  y_a^{ij} \left( u_i u_j^c h_u^a + d_i d_j^c h_d^a + \nu_i \nu_j^c h_u^a + e_i e_j^c h_d^a \right) \, ,
		\label{eq:mass-dirac}
	\end{eqnarray}
	where the color indices are suppressed for convenience. 
	
	After the EWSB, the vev $\langle h_u^a \rangle \sim v_u^a$ provides Dirac masses to up-type quarks and neutrinos, while down-type quarks and charged leptons obtain masses from the vev $\langle h_d^a \rangle \sim v_d^a$, giving the usual quark and lepton mass matrices in the PS model at tree-level:
	\begin{eqnarray}
		m_u^{ij} = m_\nu^{ij}  = y_a^{ij} v_u^a \, , \quad m_d^{ij} = m_e^{ij} =  y_a^{ij} v_d^a \, .
		\label{eq:mass-dirac2}
	\end{eqnarray}
	
	We readily observe that, couplings with bi-doublets $h^a$ give identical contributions to masses of down-quarks and masses of charged leptons for all three generations, which is apparently contradicting to known phenomenology. Also note that, if there is only one Higgs bi-doublet $h^a$ with $a=1$, the ratio of up-type quark masses to down-type quark masses will be fixed to $m_u/m_d = m_c/m_s=m_t/m_b=v_u/v_d$, which is also not consistent with the observed mass hierarchies. For simplicity, we will consider $a=2$ in the following.

	The second and the third term in eq.~(\ref{eq:Yukawa}) are expanded as:
	\begin{eqnarray}
		F_L^i H_L T  &=& 2 (u_i e_{H_L}-d_i \nu_{H_L} )  \bar{T}_3  - 2 (e_i u_{H_L}-\nu_i d_{H_L} )  \bar{T}_3 +  2  (u_i d_{H_L}  - d_i u_{H_L})T_3 \, , \nonumber \\
		\bar{F}_R^j \bar{H}_R T &=&  2 (  u_j^c e^c_{H_R} - d_j^c \nu^c_{H_R} ){T}_3 - 2 ( e_j^c u^c_{H_R} -\nu_j^c d_{H_R}^c ){T}_3 +2  (u_j^c  d_{H_R}^c - d_j^c u_{H_R}^c)\bar{T}_3 \, .
		\label{eq:mixing-expand}
	\end{eqnarray}
	
	This expansion clearly shows that only the states $d_i$ and $d_j^c$ couple to the neutral states $\nu_{H_L} $ and $\nu_{H_R}^c$, and hence, after the Higgses $H_L$ and $\bar{H}_R$ acquire vevs at their neutral directions $\langle H_L \rangle =\langle  \nu_{H_L} \rangle$ and $\langle \bar{H}_R \rangle = \langle \nu_{H_R}^c \rangle$, mixings of the following forms are generated:
	\begin{eqnarray}
		y_L^{i} F_L^i H_L T  + y_R^{j} \bar{F}_R^j \bar{H}_R T \to y_L^{i}  \langle \nu_{H_L} \rangle   (d_i \bar{T}_3)  + y_R^{j}  \langle\nu^c_{H_R} \rangle  (d_j^c  {T}_3 )\, .
	\end{eqnarray}
	
	The  sextet $T$ acquires a large mass from the operator $TT\phi$, which can then be integrated out at a high energy scale $M_T\sim \langle \phi \rangle$ by combining $\bar{T}_3$ and ${T}_3$, leaving an effective operator that combines $d_i$ and $d_j^c$ to form the massive states:
	\begin{eqnarray}
		(y_L M_T^{-1} y_R^T)^{ij}  (F_L^i H_L) (\bar{F}_R^j \bar{H}_R) \to (y_L M_T^{-1} y_R^T)^{ij}   \langle  \nu_{H_L} \rangle \langle \nu_{H_R}^c \rangle (d_i d_j^c) \, , 
		\label{eq:md-eff}
	\end{eqnarray}
	
	Therefore, down-type quarks receive extra contributions to their masses from the above effective operator by integrating out the heavy sextet $T$. Collecting together eq.~(\ref{eq:mass-dirac2}) with $a=2$ and eq.~(\ref{eq:md-eff}), the mass matrices for the quarks and leptons are now:
	\begin{eqnarray}
		m_u^{ij} &=& y_1^{ij} v_{u}^1 + y_2^{ij} v_{u}^2 \, , \nonumber\\ 
		m_d^{ij} &=& y_1^{ij} v_{d}^1 + y_2^{ij} v_{d}^2 -(y_L M_T^{-1} y_R^T)^{ij} v_L \bar{v}_R \, , \nonumber\\
		m_e^{ij} &=& y_1^{ij} v_{d}^1 + y_2^{ij} v_{d}^2  \, , \nonumber \\
		m_{\nu_D}^{ij} &=& y_1^{ij} v_{u}^{1} + y_2^{ij} v_{u}^{2}  \, .
	\end{eqnarray}
	where we have used a short-handed notation to denote the vevs of the Higgs fields as $ \langle H_L \rangle=v_L$, $ \langle \bar{H}_L \rangle=\bar{v}_L$, $ \langle H_R \rangle = v_R$ and $\langle \bar{H}_R \rangle = \bar{v}_R$.

	To break the PS symmetry down to the SM gauge group, the Higgses $H_R$ and $\bar{H}_R$ should acquire vevs at a high scale $v_{\rm PS} \gg v_{EW} $, so the following condition for vevs should be held
		\begin{eqnarray}
		v_R \sim   \bar{v}_R   \sim v_{PS} \gg v_{EW} \, .
	\end{eqnarray} 
	While the vevs $\langle h_u^a \rangle$, $\langle h_d^a \rangle$, $\langle H_L \rangle$ and $\langle \bar{H}_L \rangle$ break the $SU(2)_L$ symmetry simultaneously, their values should be bounded by the EW-scale to ensure that the $SU(2)_L$ W, Z bosons obtain the experimentally measured masses:
	\begin{eqnarray}
		v_L^2 + \bar{v}_L^2 + \sum_a \left[ (v_{u}^a)^2 + (v_{d}^a)^2 \right]  = v_{EW}^2 \, .
		\label{eq:vev-ew}
	\end{eqnarray}
	
	These relations among the vevs should be considered together when a proper scanning in the parameter space of the Yukawa couplings $(y_{a}^{ij}, y_L^{i}, y_R^{i} ,v_u^a,v_d^a,v_L, \bar{v}_R)$ is performed to fit the mass matrices in accordance with the observed quark and lepton masses and the mixing angles, similar to what has been done in~\cite{fermionfit}. In principle there are enough parameters to fit all the 19 observables of the SM while being compatible with the present neutrino data discussed in subsection \ref{sec:neutrino}.

	It is notable that, in this model, the presence of a higher dimensional effective operator $(y_L M_T^{-1} y_R^T)^{ij}  (F_L^i H_L) (\bar{F}_R^j \bar{H}_R)$ makes it possible to split the masses of down quarks and charged leptons. This differs from the usual PS models or the SO(10) model with an intermediate PS scale where the adjoint or larger representations are usually introduced to achieve the same goal. This makes the model more economical and easier to be constructed from string theory, particularly in cases where the adjoints are absent, such as the four-dimensional free fermionic constructions of heterotic models~\cite{Antoniadis:1990hb}.

	\subsection{Neutrino masses}
	\label{sec:neutrino}
	The last two terms in the superpotential of eq.~(\ref{eq:Yukawa}) are used to generate masses of neutrinos. At the true vacuum where all the vevs are acquired, the operator $\bar{F}_R H_R \psi$ produces a mixing between the right-handed neutrinos $\nu^c$ and the singlets $\psi^m$, and also the operator $\psi \psi \phi$ generates heavy Dirac masses of the singlet fields $\psi^m$:
	\begin{eqnarray}
		&y_\psi^{jm} \bar{F}_R^j H_R \psi^m  =  y_\psi^{jm}  (u_j^c \bar{u}_H^c + d_j^c \bar{d}_H^c +e_j^c\bar{e}_H^c+\nu_j^c \bar{\nu}^c_H)\psi^m \xrightarrow{\langle H_R \rangle}  y_\psi^{jm}  {v}_R ( \nu_j^c \psi^m )  \, , \nonumber \\
		& y_\phi^{mn} \psi^m \psi^n \phi  \xrightarrow{\langle \phi \rangle}   y_\phi^{mn}   v_\phi ( \psi^m \psi^n ) \,  .
	\end{eqnarray}
	
	In addition, since the $\mathbb{Z}_n$ charge of the field $\bar{H}_L$ is not specified yet, it is possible to include an additional term $\lambda_\psi^{im} F_L^i \bar{H}_L \psi^m$ in a specific model, which generates a mixing between the left-handed neutrinos and the singlets $\psi^m$ when the Higgs $\bar{H}_L$ acquires a vev at $\langle \bar{H}_L \rangle = \bar{v}_L$. The inclusion of this operator will not significantly affect the mass eigenvalues of the neutrinos as the vev $\bar{v}_L$ is expected to be much smaller than the vev $v_R$.
	\begin{eqnarray}
		\lambda_\psi^{im}  F_L^i \bar{H}_L \psi^m  =  \lambda_\psi^{im}  (Q_i \bar{Q}_{H_L} + \ell_i \bar{\ell}_{H_L} )\psi^m \xrightarrow{\langle \bar{H}_L \rangle}  \lambda_\psi^{im}  \bar{v}_L ( \nu_i \psi^m )  \, .
	\end{eqnarray}
	
	As a result, the mass matrix of neutrinos in the basis $(\nu, \nu^c, \psi)$ is given by
	\begin{eqnarray}
		\begin{pmatrix} 
			0 & m_{\nu_D}^{ij} & \lambda_\psi^{im} \bar{v}_L \\
			m_{\nu_D}^{ji} & 0 & y_\psi^{jm}  v_R  \\
			\lambda_\psi^{mi} \bar{v}_L & y_\psi^{mj}   v_R &  y_\phi^{mn}   v_\phi  \\
		\end{pmatrix} \equiv
		\begin{pmatrix} 
			0& m_u^{ij} & m_L^{im} \\
			m_u^{ji} & 0 & m_{R}^{jm} \\
			m_L^{mi} & m_{R}^{mj} & m_\psi^{mn} \\
		\end{pmatrix} \, ,
	\end{eqnarray}
	where $m_{\nu_D}^{ij}$ is replaced by $m_u^{ij}$ following eq.~(\ref{eq:mass-dirac2}), $ m_L^{im} \equiv \lambda_\psi^{im} \bar{v}_L$ , $m_R^{jm} \equiv y_\psi^{jm}   v_R$, and $m_\psi^{mn} \equiv y_\phi^{mn}   v_\phi$.
	After diagonalization, the masses of the lightest neutrinos are given by the eigenvalues
	\begin{eqnarray}
		m_{\nu} = m_u (m_R^T)^{-1} m_\psi (m_R)^{-1} m_u^T - m_L (m_R)^{-1} m_u^T - m_u (m_R^T)^{-1} m_L^T \, ,
		\label{eq:neutrinoseesaw}
	\end{eqnarray}
	where the first term is a type of contribution given by the double seesaw mechanism while the second and third terms are forms of linear seesaw~\cite{neutrinoseesaw}. 
	
	It is convenient to parametrize the ratio between the PS scale and the intermediate scale by a dimensionless ratio $r$ where $v_\phi/v_R \equiv r \ll 1 $ and thus $m_\psi = (r y_\phi /y_\psi)m_R$. Now, if the Yukawa coupling $\lambda_\psi$ is sufficiently small or if it is forbidden due to symmetry reasons, for instance by assigning certain $\mathbb{Z}_n$ charge for the field $\bar{H}_L$ so that the term $m_L $ is eliminated, the dominant contribution to the masses of light neutrinos comes from the first term $ m_u (m_R^T)^{-1} m_\psi (m_R)^{-1} m_u^T$ which is roughly $r m_u^2/m_R$. In such cases, when taking $m_u$ to be the largest value which is the top quark mass 173 GeV, for the masses of the left-handed neutrinos to be lower than the order of $10^{-2}$ eV, it is expected that $ r/m_R$ should be at the order of $10^{-15}$ ${\rm GeV}^{-1}$. Therefore, the masses of neutrinos set an upper bound on the intermediate scale $v_\phi$ by satisfying $r< 10^{-1}$ (which is in accordance with our assumption) if the Pati-Salam scale is $m_R\gsim 10^{16}$ GeV assuming that the corresponding Yukawa couplings are of the same order. As usual, the masses of light neutrinos strictly restrict the PS scale as well as the intermediate scale.
	
	On the other hand, if the term $m_L$ is not vanishing, there could be some cancellation between the double seesaw contribution and the linear seesaw contribution. It is therefore possible to lower both the PS scale and the intermediate scale by adjusting the Yukawa interactions to satisfy the bounds from neutrino masses. Furthermore, it could be possible that the Yukawa couplings are neither of order one, nor of the same order, which means that certain Yukawa hierarchies exist even at the GUT scale. In such cases, these scales can be brought down a few orders of magnitude, or even to be much lower than the GUT scale which can be accessible by future colliders. This enlarges the parameter space and makes it more flexible to fit the model with the present neutrino data regarding the neutrino masses and mixing angles.

	\subsection{Decoupling of heavy Higgs color triplet}
	\label{sec:HiggsDTS}
	The bare mass terms for vector-like Higgses $H_L$, $\bar{H}_L$, $H_R$, and $\bar{H}_R$ are absent in the superpotential eq.~(\ref{eq:superpotential}), but this will not result in a massless spectrum. Instead, they become heavy and decouple from low energy theory after the PS symmetry breaking. In this section, a detailed discussion on the masses of these Higgs fourplets will be presented and an order of magnitude estimation will be given with natural Yukawa couplings of order unity.
	
	Indeed, such a mechanism for generating heavy masses for Higgs fourplets has been discussed previously in the literature~\cite{Antoniadis:1988cm,Anastasopoulos,Leontaris:2018whz} by coupling Higgses to a sextet~\footnote{Note that in contrast to previous studies, the presence of R-symmetry in this model restricts the pairs $H_R H_R$ and $\bar{H}_R \bar{H}_R$ to couple only to the sextet $D$, but not to the sextet $T$. Therefore, $D$ and $T$ have distinct roles in the present work while their masses differ: $M_T \sim v_\phi$ while $M_D \sim v_R$, and hence, they have different phenomenological implications.}, which is exactly what has been written in $W_D$ in eq.~(\ref{eq:superpotential}). After decomposing the superfields $H_R$, $\bar{H}_R$ $D$ under the SM gauge symmetry, it is clear that the breaking of PS symmetry produces mixings between color triplet states $\bar{d}_{H_R}^c$, $d_{H_R}^c$ and $\bar{D}_3$, ${D}_3$ correspondingly. 
	\begin{eqnarray}
		\lambda_{{H}_R} H_R H_R D + \lambda_{\bar{H}_R} \bar{H}_R \bar{H}_R D \to \lambda_{{H}_R} v_R \bar{d}_{H_R}^c \bar{D}_3 + \lambda_{\bar{H}_R} \bar{v}_R d_{H_R}^c D_3  \, .
		\label{eq:mass-ctriplet}
	\end{eqnarray}
	After diagonalization, such mixing will give massive eigenstates for color triplets $\bar{d}_{H_R}^c$, $d_{H_R}^c$ by combining $\bar{D}_3$, ${D}_3$ at the PS scale, and similarly heavy states $\bar{D}_3$, ${D}_3$ with the same eigenvalue, as can be checked explicitly in eq.~(\ref{eq:mass-ctriplet3}).
	
	It is not surprising that only mixing with $d_{H_R}^c$ and $\bar{d}_{H_R}^c$ components remains, because when the $SU(4)$ gauge symmetry is broken $u_{H_R}^c$, $\bar{u}_{H_R}^c$, $e_{H_R}^c$, $\bar{e}_{H_R}^c$ fields are ``eaten'' by the Higgs mechanism, leaving only color triplets $d_{H_R}^c$  and $\bar{d}_{H_R}^c$ as ``uneaten'' fields. Combined with gauge bosons, those eaten fields will correspond to some combination of the $B-L$ generator and the $I_{3R}$ generator of the $SU(2)_R$, which also receives heavy masses at the PS scale~\cite{Anastasopoulos}. Therefore, this mechanism ensures that no light color triplets from $H_R$ and $\bar{H}_R$ can present at low energy, and thus, there will be no fast proton decay by exchanging ``uneaten'' color triplets, which will be discussed in details in section \ref{sec:proton-decay}.

	To sum up, the above mechanism is useful for naturally generating heavy masses for the Higgses $H_R$ and $\bar{H}_R$ even if a bare mass term $M_{H_R}H_R\bar{H}_R$ is absent when there is a strong global symmetry in the superpotential. From the point of view of model building, introducing these operators requires the following conditions for the $\mathbb{Z}_n$ charges:
	\begin{eqnarray}
		2n_{\bar{H}_R} +n_{D}   &=&  0 \ {\rm mod} \ n\, , \nonumber \\
		2n_{H_R} +n_D   &=&  0 \ {\rm mod} \ n \, .
		\label{eq:cc5}
	\end{eqnarray}
	Hence, to make the color triplets heavy, $H_R$ and $\bar{H}_R$ should have identical discrete charges.

	It is worth mentioning that, in addition to the dominant contribution given above, there are subdominant higher-order corrections from the non-renormalizable operators contributing to masses of $H_R$ and $\bar{H}_R$ fields. An example of such an operator is 
	\begin{eqnarray}
		\frac{c_{H_R}}{\Lambda} H_R \bar{H}_R \phi^2 \to  \frac{c_{H_R}  v_\phi^2}{\Lambda} d_{H_R}^c \bar{d}_{H_R}^c
	\end{eqnarray}
	where $c_{H_R}$ is the Wilson coefficient and $\Lambda$ can be taken to be the PS scale $v_R$. This operator contributes a mass correction $c_{H_R}  r v_\phi$ for those color triplets when $\phi$ acquires a vev at $v_\phi$, which is negligible compared to the dominant contribution that is proportional to $v_R$. On the other hand, the higher-order operator $c_D D^2 \phi^2/\Lambda$ provides a subdominant contribution to the mass of the sextet $D$. These higher-order corrections should be taken into account when their mass matrices are studied in detail.
	
	Similarly, color triplets resided in $H_L$ and $\bar{H}_L$ obtain masses via the operators
	\begin{eqnarray}
		& \lambda_{{H}_L} H_L H_L D + \lambda_{\bar{H}_L} \bar{H}_L \bar{H}_L D \to \lambda_{{H}_L} v_L {d}_{H_L} \bar{D}_3 + \lambda_{\bar{H}_L}  \bar{v}_L \bar{d}_{H_L} D_3  \, , \nonumber \\
		& \frac{c_{ H_L}}{\Lambda} H_L \bar{H}_L \phi^2 \to \frac{c_{H_L}  v_\phi^2}{\Lambda}  d_{H_L} \bar{d}_{H_L} \, .
		\label{eq:mass-ctriplet2}
	\end{eqnarray}
	
	Collecting everything mentioned above, there are three color triplets mixed with each other and under the basis $(d_{H_L}, \bar{d}_{H_R}^c, D_3)$, and their mass matrices are given by:
	\begin{eqnarray}
		\begin{pmatrix}
			c_{ H_L}r v_\phi & 0 & \lambda_{{H}_L} v_L  \\
			0 & c_{ H_R}r v_\phi & \lambda_{{H}_R} v_R   \\
			\lambda_{\bar{H}_L}  \bar{v}_L  & \lambda_{\bar{H}_R} \bar{v}_R & c_D r v_\phi \\ 
		\end{pmatrix}  \, ,
		\label{eq:mass-ctriplet3}
	\end{eqnarray}
	whose eigenvalues are approximately given by (when $r \simeq 0$):
	\begin{eqnarray}
		m_{H_L} = c_{ H_L}r v_\phi + {\cal O}(r^3)\, , \quad m_{H_R,D} = \pm \sqrt{ \lambda_{{H}_R} \lambda_{\bar{H}_R} v_R \bar{v}_R} + {\cal O}(r) \, .
	\end{eqnarray}
	Therefore, lighter color triplets in $H_L$ and $\bar{H}_L$ might reside just a few orders of magnitude below the intermediate scale $v_\phi$, while heavier color triplets in $H_R$ and $\bar{H}_R$ obtain masses at the PS scale $v_R$. 
	
	Without explicitly computing the Wilson coefficient $c_{H_L}$, we can simply estimate the result of two distinct limits, either $c_{H_L}$ is at order one so that $c_{ H_L}r v_\phi \gg \lambda_{H_L} v_L \sim \lambda_{\bar{H}_L} \bar{v}_L$, or $c_{H_L}$ is very close to zero. In the former case, the dominant contribution to masses of $H_L$ and $\bar{H}_L$ comes from higher-order corrections $c_{H_L}  r  v_\phi$ which can be much larger than the EW scale, and thus these Higgses can be integrated out and decoupled at low energy. In the latter case, the masses of color triplets in $H_L$ and $\bar{H}_L$ become comparable with the EW scale, but this will not cause any phenomenological problem since they only interact with light SM fields indirectly: they can only interact with SM fermions via exchanging the sextet $T$ or singlets $\psi^m$ as mediators. Therefore, they are decoupled in the sense that their effective couplings to SM particles are too weak as the loop process involving these fields must be suppressed by a high scale. This will become clearer when we demonstrate their effects on proton decay in section \ref{sec:proton-decay}.

	Note that in addition to the color triplets $d_{H_L}$ and $\bar{d}_{H_L}$, the other component within the supermultiplet $H_L$ and $\bar{H}_L$ also acquire heavy masses at the scale of $m_{H_L} = c_{ H_L}r v_\phi $, including the color triplets $u_{H_L}$ and $\bar{u}_{H_L}$ and the doublets $\ell_{H_L}$ and $\bar{\ell}_{H_L}$. These triplets might be indirectly probed in B physics or FCNC/LFU experiments. While a potential application to the muon anomalous magnetic moment could be considered as in Refs.~\cite{Gomez:2024dts}, our scenario operates at a large scale, where the triplets acquire substantial masses and thus contribute negligibly. One could modify the intermediate scale to accommodate possible deviations on the muon anomalous magnetic moment, but this would significantly alter other phenomenological predictions of the model. Last but not least, the proposed supersymmetric PS model indeed provides a well-motivated UV completion for the origin of triplets often used in the literature of B physics, e.g.~\cite{Buttazzo:2017ixm}.

	In conclusion, we have calculated the mixings and higher order corrections for the color triplets, and we observe that all of them can be decoupled from a low energy perspective, despite their masses being different.

	\subsection{Decoupling of heavy Higgs doublets}
	\label{sec:HiggsDoublets}
	Unlike color triplets, those electroweak doublets acting like the SM Higgs fields could be dangerous for low energy phenomenology if they are not heavy enough. There are in total four electroweak doublets within two Higgs bi-doublet $h^{a=1,2}$, meanwhile there are two more Higgs doublets $\ell_{H_L}$ in $H_L$ and $\bar{\ell}_{\bar{H}_L}$ in $\bar{H}_L$ acquiring vevs at the EW scale respectively. In total, there are three pairs of electroweak doublets with opposite hypercharges conjugating with each other. Similar to what has been done to color triplets, we need to diagonalize their mass matrix to check whether they are consistent with observed phenomenology.
	
	To give masses to these doublets and to trigger the EWSB, the supersymmetric $\mu$-term has to be included in the superpotential $W_h$~\cite{Nilles:1983ge}. We assume that the bare $\mu$-term is strictly forbidden due to higher symmetries like the conformal symmetry in the UV theory, and therefore it must be generated by quantum corrections or by the Giudice-Masiero mechanism. In our model, due to the presence of $\phi$ field, the $\mu$-term can be generated from higher order corrections via $c_h h^2 \phi^3/\Lambda^2$. This sets an upper bound for the factor $c_h r^2$:
	\begin{eqnarray}
		\frac{c_h}{\Lambda^2} h^2 \phi^3 \to c_h r^2 v_\phi h^2 \to \mu h^2 \\
		c_h r^2 \simeq v_{EW}/v_\phi \simeq 10^{-13}
	\end{eqnarray}
	Taking the $\mu$ term to be of the order of the electroweak scale, it seems that fine-tuning for the coefficient $c_h$ cannot be avoided. Similar fine-tuning issues are also discussed in~\cite{Cvetic:2015txa}.

\begin{figure}[h!]
		\centering
		\includegraphics[width=0.4\textwidth]{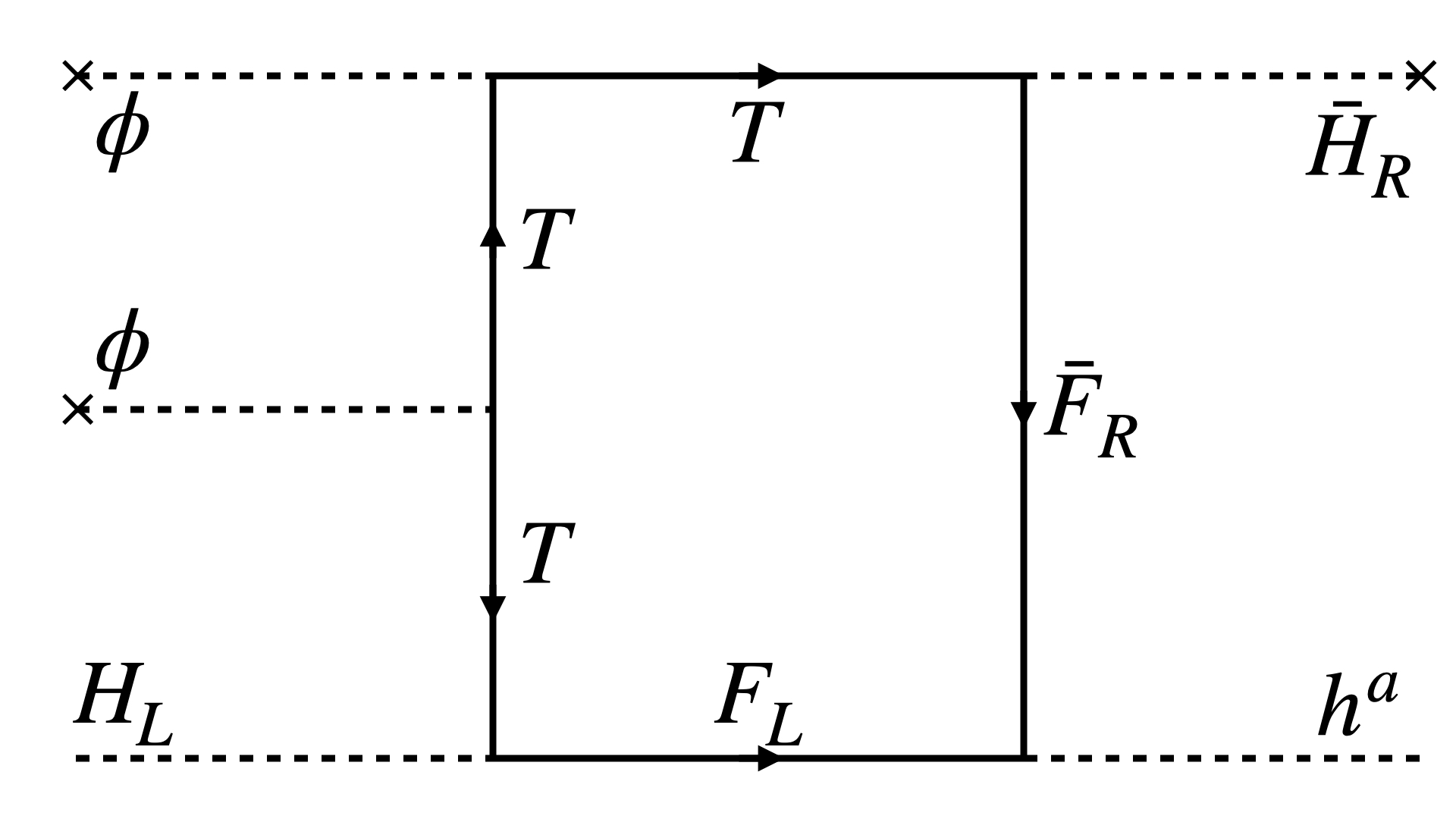}
		\quad
		\includegraphics[width=0.4\textwidth]{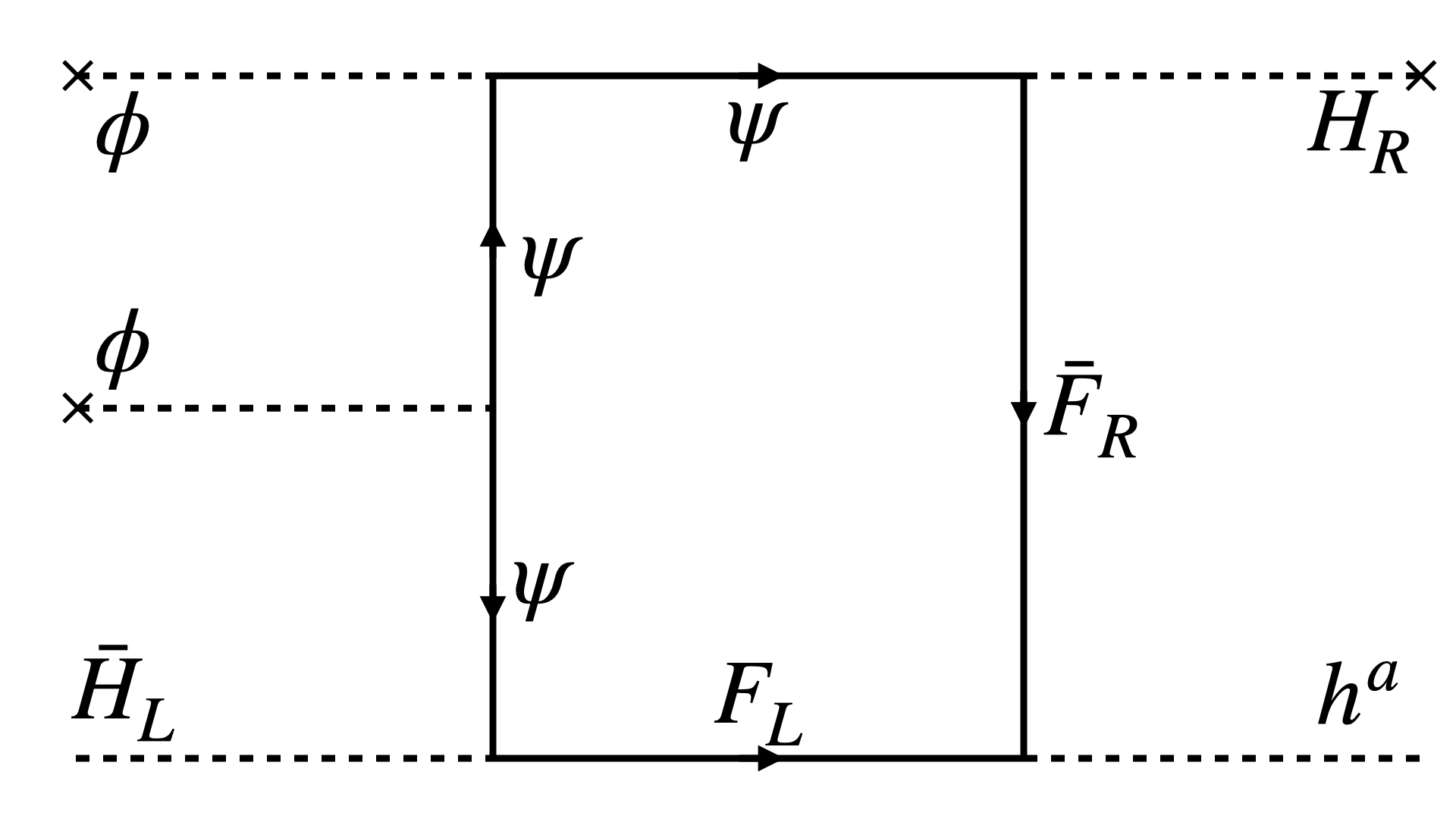}
		\caption{The shown loop process generates the higher order corrections shown in eq.~(\ref{eq:mass-doublet}).}
		\label{fig:HLHRloop}
	\end{figure}

	Apart from the $\mu$ term, there is no other mixing between these electroweak doublets at the renormalizable level. At a non-renormalizable level, the dominant contribution to the masses of Higgses doublets $\ell_{H_L}$ and $\bar{\ell}_{H_L}$ come from the operator $ c_{ H_L} H_L \bar{H}_L \phi^2/{\Lambda}  \to c_{ H_L}r v_\phi \ell_{H_L} \bar{\ell}_{H_L}$ that discussed in the previous subsection. Meanwhile, a mixing can be induced when the heavy sextet $T$ is integrated out to form an effective operator $c_{1} H_L \bar{H}_R h {\phi^2}/{\Lambda^2}$. Similarly, the effective operator $c_{2} \bar{H}_L {H}_R h {\phi^2}/{\Lambda^2}$ can be generated by integrating out the heavy singlet $\psi$. These operators can be formed by the box diagrams shown in Figure.~\ref{fig:HLHRloop}. After the PS symmetry is broken and $\phi$ develops a vev, the operator $c_{1} H_L \bar{H}_R h {\phi^2}/{\Lambda^2}$ generates mixings between the doublets  $h^{a=1,2}$ and $\ell_{H_L}$, and also the effective operator  $c_{2} \bar{H}_L {H}_R h {\phi^2}/{\Lambda^2}$ gives mixings between the doublets $h^{a=1,2}$ and $\bar{\ell}_{H_L}$:
	\begin{eqnarray}
		&\frac{c_{1}}{\Lambda^2} H_L \bar{H}_R h {\phi^2}   \to \frac{c_{1}}{\Lambda^2}  \bar{v}_R v_\phi^2 h_u^a \ell_{H_L} \equiv m_{\ell} h_u^a \ell_{H_L} \, , \nonumber \\
		&\frac{c_{2}}{\Lambda^2} \bar{H}_L {H}_R h {\phi^2}   \to \frac{c_{2}}{\Lambda^2}  {v}_R v_\phi^2 h_d^a \bar{\ell}_{H_L} \equiv m_{\bar{\ell}} h_d^a \bar{\ell}_{H_L}  \, ,
		\label{eq:mass-doublet}
	\end{eqnarray}
	where $m_\ell = c_1 v_\phi$, $m_{\bar{\ell}} = c_2 v_\phi$ when the integration scale $\Lambda$ is taken to be the PS scale.

	Therefore, the mass matrix for the three Higgs doublets under the basis $(h_d^1, h_d^2, \ell_{H_L})$ is
	\begin{eqnarray}
		\begin{pmatrix}
			c_{h1} r^2 v_\phi & c_{h12} r^2 v_\phi & c_2 r v_\phi  \\ 
			c_{h12} r^2 v_\phi & c_{h2} r^2 v_\phi & c_2 r v_\phi \\
			c_1 r v_\phi & c_1 r v_\phi &  c_{H_L} r v_\phi \\
		\end{pmatrix} \, ,
	\end{eqnarray}
	whose eigenvalues are given by
	\begin{eqnarray}
		& m_{h_1} \simeq \frac12 (c_{h1}+c_{h2} -2 c_{h12} )  r^2 v_\phi+ {\cal O}(r^3) \, , \quad \nonumber \\
		& m_{h_2, \ell_{H_L}} \simeq \pm \sqrt{2c_1 c_2} r v_\phi +   \frac14 (2 c_{H_L} + c_{h1} r +c_{h2} r + 2 c_{h12} r ) r v_\phi + {\cal O}(r^2) \, .
	\end{eqnarray}

	In conclusion, only one pair of Higgs doublets remains light at the EW scale set by the scale of $\mu$ term. The other two pairs of Higgs doublets will acquire masses near the intermediate scale $v_\phi$. This ensures that the model reduces back to the MSSM at the electroweak scale after all the heavy doublets are decoupled from the theory.

	
	\section{UV asymptotic behavior of gauge couplings}\label{sec:3}
	It is very common that when the Pati-Salam model is unified into a deeper theory such as the SO(10) grand unified theory, there is a rapid change in the RG running of gauge couplings resulting from threshold effects~\cite{Threshold} of new representations that are integrated in.

	In general, the two-loop renormalization group equations for the gauge couplings are:
	\begin{eqnarray}
		\frac{{\rm d} g_i}{{\rm d} \ln \mu} = \frac{b_i}{16\pi^2} g_i^3 + \frac{g_i^3}{(16\pi^2)^2}  \left( \sum_{j} b_{ij} g_j^2 - \sum_{\alpha} d_i^\alpha \left(y^{\alpha \dagger} y^\alpha\right)  \right) \, ,
	\end{eqnarray}
	with the one-loop and two-loop beta coefficients given by~\cite{two-loop-RGEs}
	\begin{eqnarray}
		b_i &=& -3 C_2 (G_i) +  \sum_R S (R)  \, , \nonumber \\
		b_{ij} &=& - 6 \left[C_2 (G_i)\right]^2 \delta_{ij} +  \sum_R 2 S (R) \left[  C_2 (G_i) \delta_{ij} + 2 C_2 (R)  \right]  \, ,
	\end{eqnarray}
	where $C_2 (G_i)$ is the quadratic Casimir of the adjoint representation, $S_2 (R)$ is the Dynkin index of the representation $R$ summed over all chiral multiplets. 
	
	At the symmetry-breaking scales, threshold effects are activated since heavy particles having masses at that scale start contributing to the runnings of gauge couplings, which then change the beta coefficients as well as the matching conditions of the gauge couplings at the threshold scale. At one-loop level, when the gauge group ${\cal G}$ is broken to its subgroup ${\cal H}$ at the scale $\mu$ , the matching conditions for the gauge couplings are modified by
	\begin{eqnarray}
		\alpha_{i,{\cal G}} ^{-1} (\mu )= \alpha_{i,{\cal H}} ^{-1} (\mu ) + \frac{\lambda_{i, {\cal H}} ^{\cal G}}{12 \pi} \, ,
		\label{eq:threshold}
	\end{eqnarray}
	where the one-loop matchings $\lambda_{i, {\cal H}} ^{\cal G}$ are given explicitly in~\cite{Threshold,ThresholdSO10}. In particular, neglecting the threshold corrections, when the PS symmetry is broken down to SM gauge group at the PS scale $v_R$, the matching conditions are explicitly given by
	\begin{eqnarray}
		&& 
		\alpha_{4,{\cal G}_{422}}^{-1} (M_I) = \alpha_{3,{\cal G}_{321}}^{-1} (M_I) \, , \ \nonumber \\
		&&    \alpha_{2_L,{\cal G}_{422}}^{-1} (M_I) = \alpha_{2,{\cal G}_{321}}^{-1} (M_I)  \, , \nonumber \\ 
		&& \alpha_{2_R,{\cal G}_{422}}^{-1} (M_I) =\frac53 \alpha_{1,{\cal G}_{321}}^{-1} (M_I)-\frac23 \alpha_{3,{\cal G}_{321}}^{-1} (M_I)  \, .
		\label{eq:match-gauge1}
	\end{eqnarray}

	On the other hand, the beta coefficients depend on the particle spectrum at the threshold scale, which can be specified according to the analysis in the previous section. It is most straightforward to write down the beta coefficients slightly above the EW scales, as the dynamical degrees of freedom are effectively described by the MSSM whose result has been well-studied. It is also straightforward to compute the beta coefficients above the PS scale with the complete spectrum listed in Table.~\ref{fields-Z3}. For the minimal case where there is only one copy of each sextet $N_T = N_D =1$ and two copies of Higgs bi-doublet $a=2$, we have 
	\begin{eqnarray}
		b_i = \begin{pmatrix}
			b_4 \\ b_{2_L} \\ b_{2_R}
		\end{pmatrix} = 
		\begin{pmatrix}
			0 \\ 6 \\ 6
		\end{pmatrix} \, ,
	\end{eqnarray}
	and
	\begin{eqnarray}
		b_{ij} = \begin{pmatrix}
			b_{44} & b_{42_L} & b_{42_R} \\
			b_{2_L4} & b_{2_L2_L} & b_{2_L2_R} \\
			b_{2_R4} & b_{2_R2_L} & b_{2_R2_R}
		\end{pmatrix} =
		\begin{pmatrix}
			95 & 15 & 15 \\
			75 & 60 & 6 \\
			75 & 6 & 60
		\end{pmatrix} \, .
	\end{eqnarray}
	The running behavior of $SU(2)_L$ and $SU(2)_R$ couplings looks the same because their beta coefficients preserve parity, but they get slightly modified by threshold effects when heavy particles are integrated out at the corresponding threshold scales because the spectrum is not exactly left-right symmetric. Therefore, there is enough parameter space for adjusting threshold corrections to achieve the unification of gauge couplings at the UV scale.

\begin{figure}[h!]
		\centering
		\includegraphics[width=0.6\textwidth]{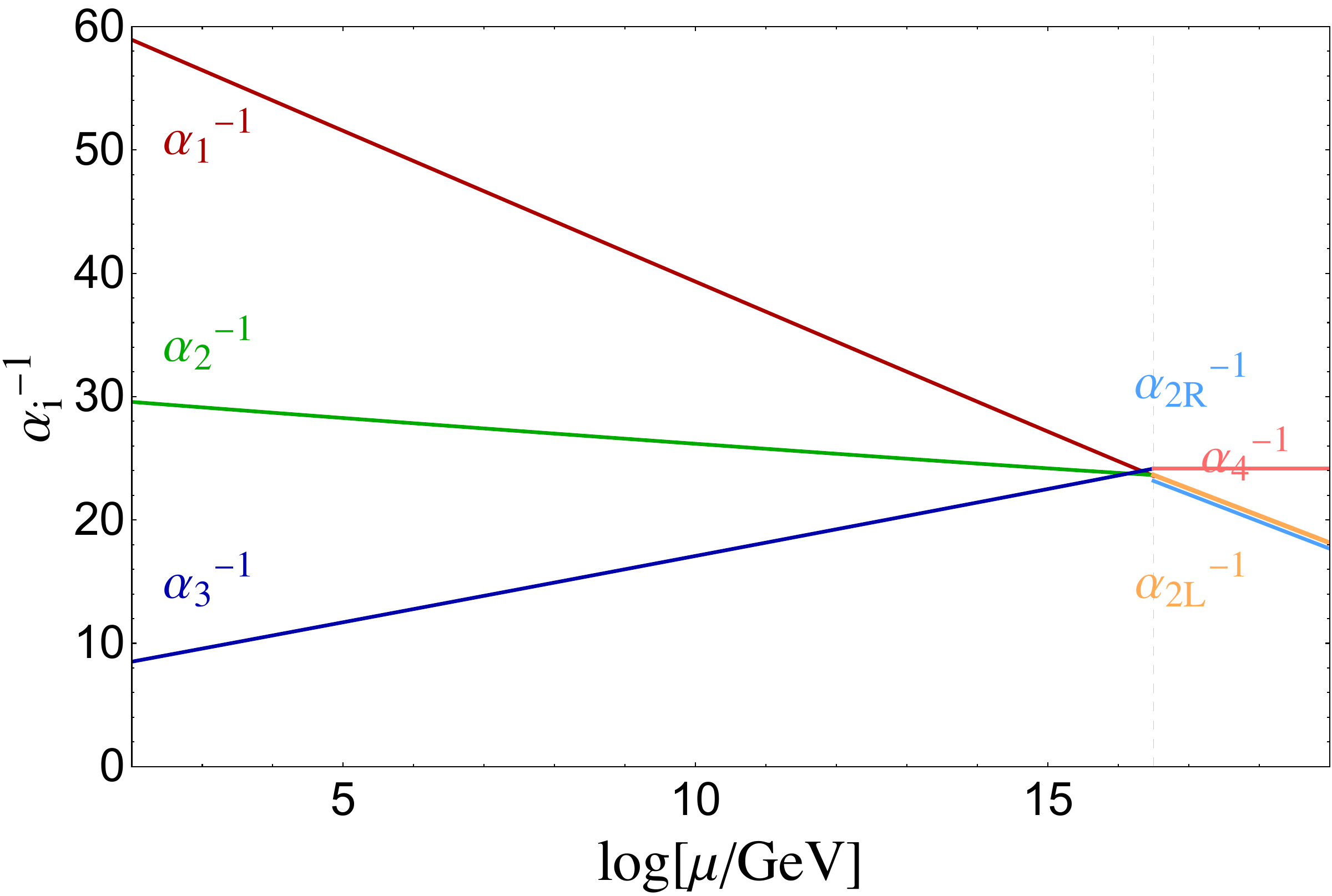}
		\caption{The RG running of the gauge couplings without having a SO(10) unification.}
		\label{fig:RG}
	\end{figure}

	As before, taking the PS scale to be roughly $10^{16}$ GeV, gauge couplings can intersect at that scale and if unification indeed exists, the PS gauge couplings stop running and are unified into a universal SO(10) gauge coupling immediately. However, if there is no unification, the $SU(4)_C$ gauge couplings become conformal at the one-loop level, while the $SU(2)_L$ and $SU(2)_R$ gauge couplings continue to increase until the Planck scale. The RG evolution of gauge couplings from the electroweak scale without having an SO(10) unification is shown in Figure ~\ref{fig:RG}.

	The initial conditions on the MSSM gauge couplings as
	($\alpha^{-1}_{1_{\rm }}, \alpha^{-1}_{ 2_{\rm }},\alpha^{-1}_{3_{\rm }}$), evaluated in the $\overline{\text{MS}}$ renormalization scheme with two-loop accuracy, are the coupling values at the electroweak scale that we take to be the $Z$ boson mass $M_Z=91.2$ GeV, namely~\cite{PDG},
	\begin{eqnarray}
		\Big( \alpha^{-1}_{1_{\rm EW}}, \  \alpha^{-1}_{2_{\rm EW}}, \  \alpha^{-1}_{3_{\rm EW}}\Big) = \Big( 59.0272 , \ 29.5879, \ 8.4678 \Big) \, ,
		\label{eq:gis}
	\end{eqnarray}
	where the hypercharge coupling $\alpha_Y$ has been normalized with the usual GUT condition leading to $\alpha_1 / \alpha_Y=5/3 $.  In the equation above, we have neglected for convenience the experimental errors on the inverse coupling constants (as well as the estimated theoretical uncertainties) and kept only the central values. These errors, in particular the one that affects the strong coupling $\alpha_3$, will lead to an uncertainty on the obtained scales $M_U$ and $M_I$ of the order of a few percent at most and will therefore not affect our discussion in a significant way. 
	
	It is noted that there will be no Landau pole from the PS scale to the Planck scale. If the model can be ultimately unified into a string theory or a quantum theory of gravity, there could be large threshold corrections near the quantum gravity scale, such as the KK towers from compactification of extra dimensions or string towers. Without string thresholds, the gauge couplings are safe and the model is still consistent at the far UV scale.

	\section{Proton decay}
	\label{sec:proton-decay}
	The Pati-Salam gauge symmetry $SU(4)_C \times SU(2)_L \times SU(2)_R$ inherently conserves $U(1)_{B-L}$ at the renormalizable level. Consequently, the $SU(4)_C$ gauge bosons, which mediate transitions between quarks and leptons, do not induce proton decay in the minimal PS constructions.  This property allows for lowering the PS symmetry breaking scale without conflicting with experimental bounds~\cite{Dutka:2022lug}. However, at the non-renormalizable level, baryon-number violating operators emerge from integrating out heavy states, necessitating a careful analysis of the various proton decay channels.
	
\begin{figure}[h!]
		\centering
		\includegraphics[width=0.3\textwidth]{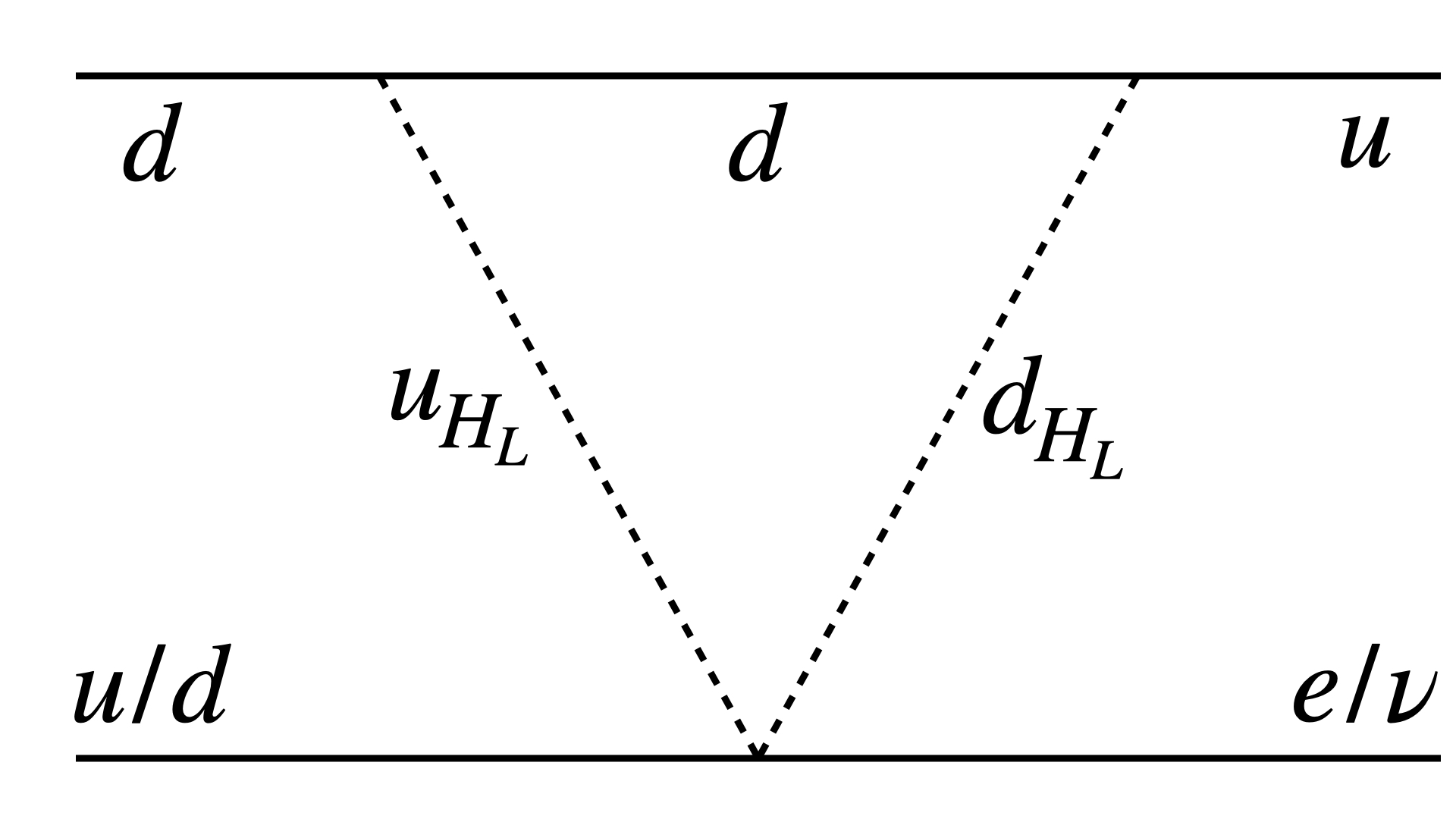}
		\qquad
		\includegraphics[width=0.3\textwidth]{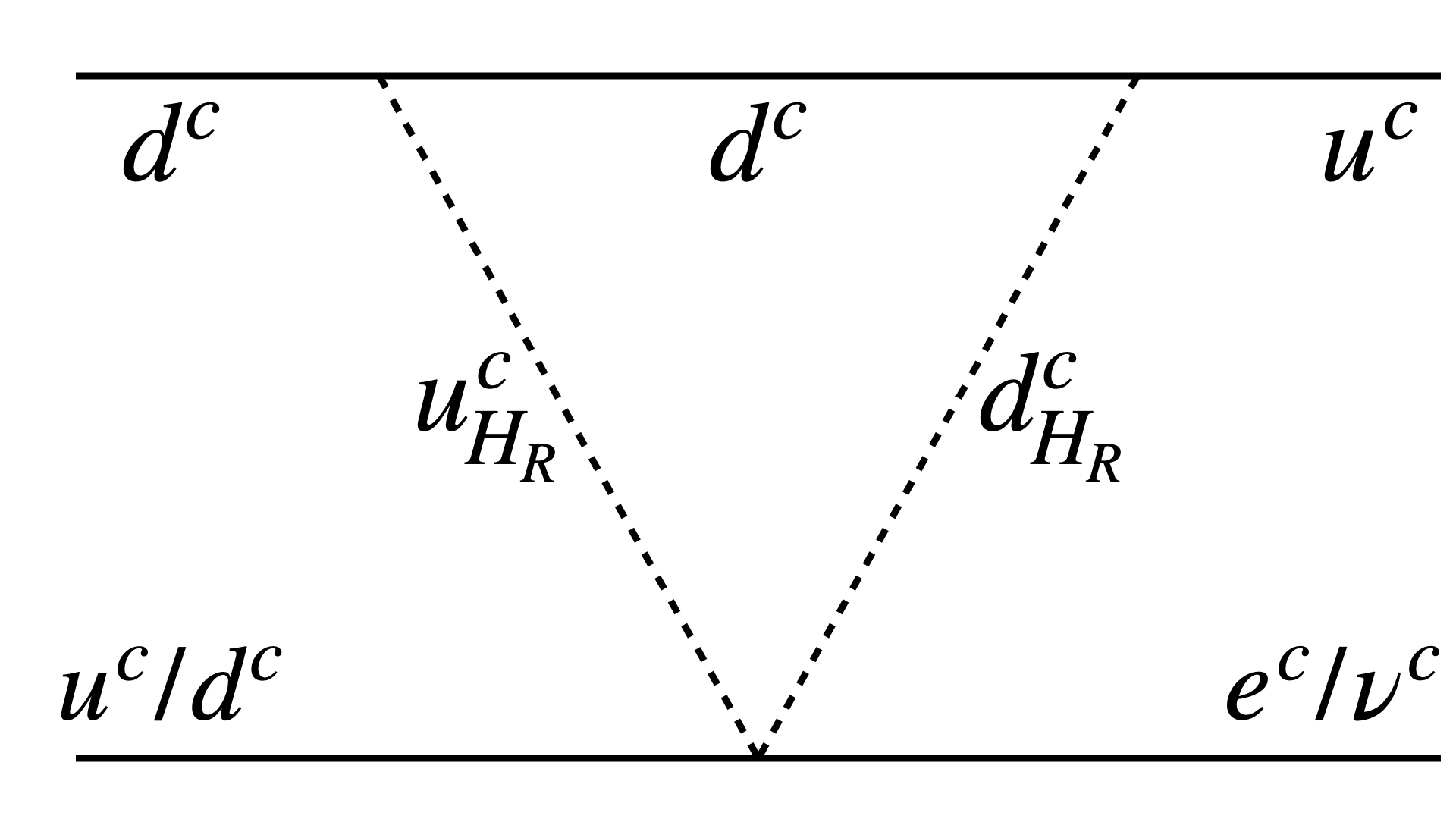}
		\caption{Diagrams associated with the proton decay operators in eq.~(\ref{PD02}). They are not symmetric because of the vast mass difference between the scalar mediator within the superfields $u_{H_L}$, $d_{H_L}$, and those within the superfields $u_{H_R}^c$, $d_{H_R}^c$.}
		\label{fig:protondecay}
	\end{figure}

	In the proposed model, the dominant baryon-number violating operators with $\Delta B = 1$ arises from the superpotential couplings with the heavy sextets $T^k$ in eq.~(\ref{eq:Yukawa}). Integrating out $T^k$ generates the dimension-5 effective operators $(F_L H_L) (F_L H_L)$ and $(\bar{F}_R \bar{H}_R )(\bar{F}_R \bar{H}_R )$ that violate the baryon number conservation. Similar to eq.~(\ref{eq:mixing-expand}), the PS multiplets in these operators can be expanded into SM components:
	\begin{equation}
        \label{PD01}
        \begin{split}
		\frac{y_L^2}{M_T} (F_L H_L ) (F_L H_L )  &\supset  \frac{y_L^2}{M_T} (u d_{H_L}  - d u_{H_L})\left[  (  u e_{H_L} - d \nu_{H_L} )-  ( e u_{H_L} -\nu d_{H_L} )\right] \, ,  \\
		\frac{y_R^2}{M_T}(\bar{F}_R \bar{H}_R ) (\bar{F}_R \bar{H}_R )&\supset  \frac{y_R^2}{M_T}   (u^c d_{H_R}^c-d^c u_{H_R}^c ) \left[  (u^c  e_{H_R}^c - d^c \nu_{H_R}^c) - (e^c u^c_{H_R} -\nu^c d_{H_R}^c ) \right] \, .
        \end{split}
	\end{equation}
    These are the only baryon-number violating operators at dimension-5, since only the combination $(F_L H_L ) (F_L H_L )$, $(\bar{F}_R \bar{H}_R ) (\bar{F}_R \bar{H}_R )$, and $(F_L H_L ) (\bar{F}_R \bar{H}_R )$ have non-vanishing CG coefficients. 
	Replacing the $H_L$ and $\bar{H}_R$ with their vevs, these operators are explicitly:
	\begin{equation}
        \label{PD02}
        \begin{split}
		{\cal O}^{\Delta B =1}  &=  \frac{y_L^2}{M_T} \left[(v_L +\nu_{H_L}) (ddu_{H_L} - ud d_{H_L})- (ue+\nu d)u_{H_L} d_{H_L} \, \right.  \\
		  & \left. + \  u\nu d_{H_L} d_{H_L} + de u_{H_L} u_{H_L} + uu d_{H_L} e_{H_L} - ud u_{H_L} e_{H_L} \right] \, ,  \\
		{\cal O}^{\Delta B =-1} & =   \frac{y_R^2}{M_T} \left[ (\bar{v}_R+\nu_{H_R}^c)(d^c d^c u^c_{H_R} -u^c d^c d^c_{H_R}) - (u^c e^c+\nu^c d^c)u^c_{H_R} d^c_{H_R} \, \right.  \\
		  & \left. + \ u^c\nu^c d^c_{H_R} d^c_{H_R} + d^c e^c u^c_{H_R} u^c_{H_R} + u^c u^c d^c_{H_R} e^c_{H_R} - u^c d^c u^c_{H_R} e^c_{H_R} \right] \, .
        \end{split}
	\end{equation}
	The first term mediates the proton decay process through diagrams involving virtual color triplets shown in Figure.~\ref{fig:protondecay}, which is suppressed by a loop factor and the mass of the heavy sextet $T$. However, if the intermediate scale $v_\phi$ is not as high as the GUT scale, the proton decay amplitude can be enhanced. The proton lifetime thus puts a constraint on the lower bound of the intermediate scale. 

	The loop diagrams in Figure.~\ref{fig:protondecay} can be evaluated using the Passarino-Veltman method~\cite{PassarinoVeltman}. The final amplitude is proportional to a loop factor $F(m,m_1^2,m_2^2)$ given in ref.~\cite{Ellis:2019fwf}:
	\begin{eqnarray}
		F(m,m_1^2,m_2^2) &\equiv & \int \frac{d^4 k}{\pi^2} \frac{i}{(\slashed{k} - m ){(k^2 - m_{1}^2)} ({k^2- m_2^2 })}    \nonumber \\
		&=&  \frac{m}{m_1^2 - m_2^2} \left[ \frac{m_1^2}{m_1^2 - m^2} \ln\left( \frac{m_1^2}{m^2}\right) - \frac{ m_2^2 }{m_2^2-m^2}\ln\left( \frac{m_2^2}{m^2}\right) \right] \nonumber \\
		&  \xrightarrow{m\ll m_{1,2}}&  \frac{m}{m_1^2 - m_2^2}  \ln\left( \frac{m_1^2}{m_2^2}\right)   \, .
	\end{eqnarray}
	Therefore, the amplitude of the left figure is proportional to:
	\begin{eqnarray}
		\frac{y_L^6 v_L^2}{16\pi^2 M_T^3} \cdot F(m_d,m_{u_{H_L}}^2,m_{d_{H_L}}^2) &=& \frac{y_L^6 v_L^2}{16\pi^2 M_T^3}  \frac{m_d}{m_{u_{H_L}}^2 - m_{d_{H_L}}^2}  \ln\left( \frac{m_{u_{H_L}}^2}{m_{d_{H_L}}^2}\right) \nonumber \\
		&\sim& 10^{-50} \frac{1}{m_{u_{H_L}}^2 - m_{d_{H_L}}^2}  \ln\left( \frac{m_{u_{H_L}}^2}{m_{d_{H_L}}^2}\right)  \, ,
	\end{eqnarray}
	and similarly, the amplitude of the right one is proportional to:
	\begin{eqnarray}
		\frac{y_R^6 v_R^2}{16\pi^2 M_T^3} \cdot F(m_{d^c},m_{d_{H_R}^c}^2,m_{u_{H_L}^c}^2)&=& \frac{y_R^6 v_R^2}{16\pi^2 M_T^3}  \frac{m_d}{m_{d_{H_R}}^2 - m_{u_{H_R}}^2}  \ln\left( \frac{m_{u_{H_R}}^2}{m_{d_{H_R}}^2}\right) \nonumber \\
		&\sim& 10^{-20} \frac{1}{m_{u_{H_R}}^2 - m_{d_{H_R}}^2}  \ln\left( \frac{m_{u_{H_R}}^2}{m_{d_{H_R}}^2}\right)  
		\, .
	\end{eqnarray}
	where we have taken $m_d \lesssim 4$ GeV, $y_L \sim y_R \sim 0.1$, the mass of sextets at $M_T \sim 10^{15}$ GeV, the PS scale at $ v_R \sim \Lambda \sim 10^{16}$ GeV, and $v_L \sim 10^2$ GeV for a rough estimation of their magnitude. The mass difference $m^2_{u_{H_R}}-m^2_{d_{H_R}} $ should lie between the range of ${\cal O}(10^{-2}) v_R^2 \sim {\cal O}(10^2) v_R^2$, approximately at the GUT scale. The mass difference $m^2_{u_{H_L}}-m^2_{d_{H_L}} $, unless they are extremely light (less than 1 GeV) which is hardly possible with the present analysis, further suppresses the amplitude. Thus, we can readily observe that the amplitudes of the two diagrams are naturally suppressed by a cubic power of the GUT scale, and are much smaller than the current experimental bound on the proton decay amplitude. 

    The suppression for proton decay in the proposed model is not surprising, after all, the PS gauge symmetry and the supersymmetry are both very restrictive in limiting the possible baryon-number violating operators. A similar mechanism that suppresses proton decay mediated by the color-triplet Higgsino is also observed in SUSY-$SO(10)$~\cite{Babu:1993we}, where a doublet–triplet splitting by the adjoint Higgs converts those dimension 5 operators effectively into dimension 6, resulting in a natural suppression of proton decay. The difference is that, in our approach, the color triplets become massive due to the vevs of bifundamentals instead of the adjoint.

    \section{Cosmological implications}\label{sec:5}
   The spontaneous breaking of discrete or continuous symmetries in the early universe can lead to the formation of topological defects, such as monopoles, cosmic strings, or domain walls, depending on the symmetry group and breaking pattern~\cite{defects}. In this model, the PS symmetry breaking at the GUT scale ($v_R \sim 10^{16}$ GeV) would generate superheavy magnetic monopoles~\cite{Preskill:1979zi}, while the subsequent breaking of the discrete $\mathbb{Z}_3$ symmetry at an intermediate scale ($v_\phi$) produces domain walls. However, if these defects form before or during inflation, their energy density is exponentially diluted by the inflationary expansion, rendering them unobservable~\cite{Linde:1990flp}. To preserve observable signatures, such as gravitational waves (GWs) from collapsing domain walls, in the present study we assume that the $\mathbb{Z}_3$-breaking scale is of the order of $10^{11}$ GeV.
	In this section, we explore how the interplay of symmetry-breaking scales in our model could yield detectable cosmological signals, focusing on domain walls and their GW signatures.

	To reconcile the $\mathbb{Z}_3$-breaking scale $v_\phi \sim 10^{11}$ GeV with neutrino mass constraints, we need to relax the tight correlation between the PS scale $v_R$ and $v_\phi$ inherent in the seesaw mechanism in eq.~(\ref{eq:neutrinoseesaw}) due to assuming the double seesaw term dominates. Indeed, in our neutrino seesaw framework, the linear seesaw contribution allows for a lower $v_\phi$ if cancellations occur between the double and linear seesaw terms. We can thus retain $v_R \sim 10^{16}$ GeV and set $v_\phi \sim 10^{11}$ GeV by assuming certain tuning on the model parameters in eq.~(\ref{eq:neutrinoseesaw}), ensuring that domain walls form in a post-inflation epoch while avoiding the problem of monopoles overproduction.
	
	At energies well below $v_\phi$, the heavy PS degrees of freedom such as the Higgses $H_R$, $\bar{H}_R$ and the sextet $D$ decouple, leaving an effective theory described by the MSSM supplemented by gauge singlets $\phi$, etc. The renormalizable superpotential for $\phi$ is: 
	\begin{eqnarray}
		W_{\phi} = y_{\phi}^{mn} \psi^m \psi^n  \phi  + y_{T} T T \phi + \lambda_\phi \phi^3\, ,
	\end{eqnarray}
    supplemented by a soft supersymmetry breaking scalar mass term $m_\phi^2 |\phi|^2$. The origin of those soft supersymmetry breaking terms might be, depending on the specific breaking mechanism, obtained from a hidden sector that is coupled to the visible sector by either gauge~\cite{Gaugemediate} or gravitational interactions~\cite{Gravitymediate}.
    
    At lower energy, after integrating out the F-term, the corresponding renormalizable potential for $\phi$ is~\cite{Ellis:1986mq}
	\begin{eqnarray}
		V(\phi) = m_\phi^2 |\phi|^2 + (A m_\phi \lambda_\phi \phi^3 + {\rm h.c.}) + |\lambda_\phi \phi^2|^2 \, ,
	\end{eqnarray}
	where $A$ is a numerical coefficient and $|A|>1$ which has to be determined by solving the full action involving the K\"ahler potential. Taking $\lambda_\phi$ real and parameterizing $\phi = v_0 e^{i \varphi}$, the potential simplifies to:
	\begin{eqnarray}
		V(v_0,\varphi)= m_\phi^2 v_0^2 + 2A m_\phi \lambda \ \cos 3\varphi \ v_0^3 + \lambda_\phi^2 v_0^4 \, ,
	\end{eqnarray}
	which exhibits three degenerate minima at 
	\begin{eqnarray}
		v_0= \frac{m_\phi}{4\lambda_\phi} \left(3 |A| +\sqrt{9A^2-8}\right)\, , \quad \, \varphi=0,\pm\frac{2\pi}{3} \, \, (A\lambda_\phi < 0) \, .
	\end{eqnarray}
    
    Therefore, to ensure a symmetry-breaking scale at $v_0=v_\phi \sim 10^{11}$ GeV, a possible choice for the soft SUSY-breaking mass $m_\phi$ and the coefficient $A$ can be $m_\phi \gsim 10^5$ GeV and $A\gsim 10$, respectively, while the self-coupling is constrained to $\lambda_\phi \sim 10^{-5}$. For $|A|\gsim 1$, the domain wall tension $\sigma$ can be well approximated as~\cite{Wu:2022stu}
    \begin{eqnarray}
		\sigma \simeq  v_0^2  \sqrt{(-m_\phi^2 +2\lambda_\phi^2 v_0^2)} \, .
	\end{eqnarray}

 However, an exactly $\mathbb{Z}_3$-symmetric potential produces long-lived domain walls that dominate the energy density in the early universe which drives an accelerating expansion that is inconsistent with current observations~\cite{Zeldovich:1974uw}. A common way to avoid this problem is to introduce an explicit $\mathbb{Z}_3$-breaking term so that the domain walls start collapsing, which can be conveniently parametrized as~\cite{Wu:2022stu}:
	\begin{eqnarray}
		V_{\slashed{\mathbb{Z}}_3} = \frac{2e^{i \alpha}}{3\sqrt{3}}\epsilon \phi \left( \frac14 \phi^3 - v_0^3\right) + {\rm h.c.} \, ,
	\end{eqnarray}
	where $\epsilon \ll 1$ parametrizes the breaking strength, and $\alpha$ is a free parameter. This term, which could originate from Planck-suppressed operators or SUSY-breaking effects~\cite{Kamionkowski:1992mf}, lifts the degeneracy of the three vacua, creating energy differences:
    \begin{eqnarray}
		 \left( V_{\rm bias} \right)_{10} = \epsilon v_0^4 \cos (\alpha +\pi/6) \, , \quad 
		 \left( V_{\rm bias} \right)_{20} = \epsilon v_0^4 \cos (\alpha -\pi/6) \, ,
	\end{eqnarray}
    where $\left( V_{\rm bias} \right)_{ij}$ denotes the energy difference between the $i$-th and $j$-th vacua.

	The domain walls collapse~\cite{CollapsingDW} when the vacuum pressure caused by the energy bias dominates over their tension-driven surface pressure $p_T \sim {\cal A} \sigma/t$. the resulting gravitational wave spectra are characterized by a peak frequency $f_{\rm peak}$ and energy density $\Omega_{\rm GW}$~\cite{Wu:2022stu}:
	\begin{eqnarray}
		f_{\rm peak} = 1.1\times 10^{-7} {\rm Hz} \times \left( \frac{g_*(T_{\rm ann})}{10}\right)^{1/2}  \left( \frac{10}{g_{*S}(T_{\rm ann})}\right)^{1/3}  \left( \frac{T_{\rm ann}}{ {\rm GeV}}\right) \, ,
	\end{eqnarray}
	\begin{eqnarray}
		\Omega_{\rm GW}(f_{\rm peak}) h^2 = 7.2 \times 10^{-26} \times \tilde{\epsilon}_{\rm GW} {\cal A}^2 \times \left( \frac{10}{g_*(T_{\rm ann})}\right)^{4/3}  \left( \frac{\sigma^{1/3}}{\rm TeV}\right)^{6}  \left( \frac{  {\rm GeV}}{T_{\rm ann}}\right)^4 \, ,
	\end{eqnarray}
	where $T_{\rm ann}$ is the temperature at wall annihilation:
    \begin{eqnarray}
		T_{\rm ann} =3.41 \times 10^{4} \ {\rm GeV} \times C_{\rm ann}^{-1/2} {\cal A}^{-1/2} \times \left( \frac{10}{g_*(T_{\rm ann})}\right)^{1/4}  \left( \frac{\rm TeV}{\sigma^{1/3}}\right)^{3/2}  \left( \frac{V_{\rm bias}^{1/4}}{ {\rm GeV}}\right)^2   \, ,
	\end{eqnarray}
    with the parameters $\alpha = 2\pi/9$, ${\cal A} = 1.10\pm 0.20$, $C_{\rm ann}=5.02 \pm 0.44$, and $\tilde{\epsilon}_{\rm GW} \simeq 0.7 \pm 0.4$ as given in~\cite{Wu:2022stu}. The GW spectrum is approximately $\Omega_{\rm GW} h^2 \propto f^3$ below $f_{\rm peak}$ and $\propto f^{-1}$ above it~\cite{GW}, though some recent studies show deviations at high frequencies~\cite{Ferreira}.

	To quantify the gravitational wave signatures of the collapsing $\mathbb{Z}_3$ domain walls, we define two benchmark models anchored in the symmetry-breaking scale $v_\phi = 10^{11}$ GeV and the SUSY-breaking parameter $A=10$. These benchmarks are chosen to span distinct observational regimes for future GW detectors.

	\begin{figure}[t!]
		\centering
		\includegraphics[width=0.8\textwidth]{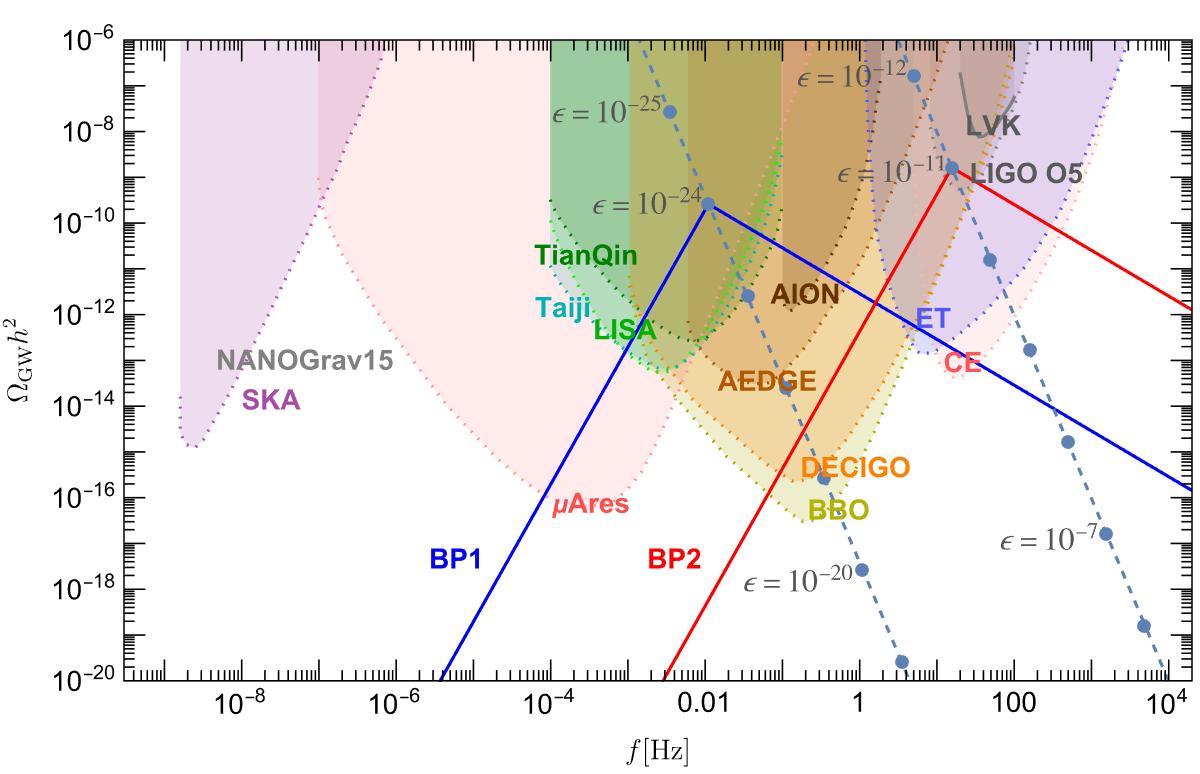}
		\caption{Predicted gravitational wave spectrum from collapsing of ${\mathbb Z}_3$ domain wall. The peak frequencies of the resulting GWs are distributed between the two dashed lines, where the soft supersymmetry breaking mass $m_\phi$ takes values ranging from $10^4 $ GeV to $ 5\times10^{11}$ GeV.
		The solid curves represent the GW background produced by the annihilation of domain walls formed after $\mathbb{Z}_3$ symmetry breaking at $v_\phi = 10^{11}$ GeV, for the biases $\epsilon = 10^{-24}$ and $\epsilon =10^{-11}$, respectively. The blue curves correspond to BP1 which peaks at approximately $10^{-2}$ Hz, while the red curves represent BP2 with a peak around $15$ Hz. Other points along the dashed lines correspond to peak frequencies for a different bias $\epsilon$.}
		\label{fig:GWs}
	\end{figure}

	The Benchmark Model 1 (BP1) assumes an explicit $\mathbb{Z}_3$-breaking bias $\epsilon = 10^{-24}$, leading to domain wall annihilation at $T_{\rm ann}=68.2 $ TeV. The resultant GW spectrum is shown in the red curve in fig.~\ref{fig:GWs}, which peaks at $f_{\rm peak} = 1.1\times 10^{-2} $ Hz with the energy density $\Omega_{\rm GW} = 2.6 \times 10^{-10}$ and fits the band targeted by space-based interferometers like LISA, TianQin, and Taiji~\cite{LISA:2024hlh,Hu:2017mde,TianQin:2015yph}. The Benchmark Model 2 (BP2) has $\epsilon = 10^{-11}$ and $T_{\rm ann} =9.6\times 10^7$ GeV, to shift the peak to $f_{\rm peak} =15.6$ Hz with the energy density $\Omega_{\rm GW} = 1.6 \times 10^{-9}$, as shown in the blue curve in fig.~\ref{fig:GWs}, within the sensitivity band of ground-based detectors like LIGO~\cite{LIGOScientific:2021nrg}. Furthermore, in the present analysis, we do not expect the produced Z3-domain wall signals to explain the PTAs, as the PTA preferred range, the nHz band, requires a much lower $\mathbb{Z}_3$-breaking scale that is inconsistent with the observed low-energy phenomenology. 

    It is worth mentioning that, while the minimal $\mathbb{Z}_3$ model provides a tractable framework for gravitational wave production, the analysis can be extended to a more general model based on $\mathbb{Z}_n$ symmetry discussed in section~\ref{sec:2}. See, for instance,~\cite {Wu:2022tpe}, for further discussions.

	\section{Conclusions}	\label{sec:6}
	In this work, based on longitudinal studies as well as recent developments in string model building, we presented a revamped version of a minimal supersymmetric Pati-Salam (PS) model as a possible improved candidate for physics beyond the Standard Model. The model features a  Higgs sector accommodated in bifundamentals $H_R+\bar H_R=(4,1,2)+(\bar 4,1,2)$, $H_L+\bar H_L=(4,2,1)+(\bar 4,2,1)$ as well as a pair of $h_a=(1,2,2), a=1,2$ where the latter produces the MSSM Higgs fields after the PS symmetry breaking. The PS symmetry is spontaneously broken near the Grand Unification scale where only the SM Higgs bosons and fermions stay light, thus below the PS breaking scale the dynamics is effectively described by the conventional MSSM which is assumed to be effective from a few orders of magnitude higher than the electroweak scale. It is also assumed that a global R-symmetry and a discrete symmetry of $\mathbb{Z}_n$, inspired from string theory constructions, are present to constrain the superpotential couplings. As an illustration, in the present work, we discussed a minimal scenario with a $\mathbb{Z}_3$ symmetry, and fixed the number of sextets to $N_T=N_D =1$. Furthermore, we have discussed in detail how such extraneous superfields are decoupled from the low-energy theory. We examined the renormalization group evolution and showed that the model is safe from any Landau poles up to the Planck scale. Proton decay has been discussed and found to be suppressed resulting to a proton lifetime beyond the present experimental bounds. 
	
	A major drawback in GUT models with Higgs fields accommodated only in small representations is that they fail to reproduce the Georgi–Jarlskog mass relations due to the fact that quarks and lepton mass matrices are restricted by GUT relations. We demonstrated how the specific Higgs sector of the proposed PS model is capable of disentangling quark and lepton masses,  thus paving the way for acceptable relations between quarks and leptons.  More precisely, it is notable that, in this model, the presence of higher dimensional effective operators $(y_L M_T^{-1} y_R^T)^{ij}  (F_L^i H_L) (\bar{F}_R^j \bar{H}_R)$ make it possible to split the masses of down quarks and charged leptons. This is different from the usual PS models or the SO(10) model with an intermediate PS scale where usually large representations or adjoints are introduced to achieve the same goal. This makes the model more economical and easier to be constructed from string theory, particularly in heterotic models where the adjoints are absent.

	The spectrum of the model includes $SU(4)_C$-sextet fields decomposed to triplet and antitriplet pairs after the PS-symmetry breaking. All but one of these pairs form Yukawa couplings with the bifundamental Higgses and decouple from the light spectrum. The leftover light color triplet-pair receives intermediate mass and weakly interacts with SM fermions, with insignificant implications in B-physics beyond present experimental sensitivity.

 In conclusion, we consider that the present analysis indicates a promising Pati-Salam model embeddable in a string theory framework that deserves further exploration. Detailed investigations of the phenomenological implications and the fermion mass matrix structure are left for future work in the context of the string-derived model.

	\bigskip 
	
	\noindent {\bf Acknowledgements:} \smallskip
	
	\noindent This work was supported by the National Natural Science Foundation of China (NSFC) under Grant Nos. 12205064, 12347103, and Zhejiang Provincial Natural Science Foundation of China under Grant No. LDQ24A050002.
	
	\setcounter{equation}{0}
	\renewcommand{\theequation}{A.\arabic{equation}}
	\setcounter{table}{0}
	\renewcommand{\thetable}{A.\arabic{table}}


	
	
	
	
\end{document}